\documentclass{jaa}
\usepackage{natbib}

\usepackage{graphicx}
\usepackage{amsmath}	
\usepackage{soul,xcolor}
\usepackage{subcaption}
\usepackage{multirow}
\usepackage{bm}
\usepackage{tikz}
\usetikzlibrary{shapes,arrows}
\usepackage{url}

\tikzstyle{startstop} = [rectangle, rounded corners, minimum width=3cm, minimum height=1cm,text centered, draw=black, fill=red!30]
\tikzstyle{io} = [trapezium, trapezium left angle=70, trapezium right angle=110, minimum width=3cm, minimum height=1cm, text centered, draw=black, fill=blue!30]
\tikzstyle{process} = [rectangle, minimum width=3cm, minimum height=1cm, text centered, draw=black, fill=orange!30]
\tikzstyle{decision} = [diamond, minimum width=3cm, minimum height=1cm, text centered, draw=black, fill=green!30]
\tikzstyle{arrow} = [thick,->,>=stealth]

\graphicspath{{./}{figures/}}
\begin{document}\sloppy

\title{Quantifying Period Uncertainty in X-ray Pulsars with Poisson-Limited Data}


\author{Akshat Singhal\textsuperscript{1, *}, Ishan Jain\textsuperscript{2, **}, Suman Bala\textsuperscript{3} and Varun Bhalerao}
\affilOne{\textsuperscript{1}Homi Bhabha Center for Science Education, Tata Institute of Fundamental Research, Mumbai - 400088, India\\}
\affilTwo{\textsuperscript{2}Department of Biomedical Engineering, Johns Hopkins University, Baltimore, MD 21218, United States\\}
\affilThree{\textsuperscript{3}Department of Physics, Indian Institute of Technology Bombay, Powai, Mumbai - 400076, India}


\twocolumn[{

\maketitle

\corres{akshat.singhal014@gmail.com\\ ** Equal contributor}

\msinfo{12 November 2023}{XX; pre-print}

\begin{abstract}
There have been significant developments in the period estimation tools and methods for analysing high energy pulsars in the past few decades. However, these tools lack well-standardised methods for calculating uncertainties in period estimation and other recovered parameters for Poisson--dominated data. Error estimation is important for assigning confidence intervals to the models we study, but due to their high computational cost, errors in the pulsar periods were largely ignored in the past. Furthermore, existing literature has often employed semi-analytical techniques that lack rigorous mathematical foundations or exhibit a predominant emphasis on the analysis of white noise and time series data. We present results from our numerical and analytical study of the error distribution of the recovered parameters of high energy pulsar data using the $Z_n^2$ method. We comprehensively formalise the measure of error for the generic pulsar period with much higher reliability than some common methods. Our error estimation method becomes more reliable and robust when observing pulsars for few kilo-seconds, especially for typical pulsars with periods ranging from a few milliseconds to a few seconds. 
We have verified our results with observations of the \emph{Crab} pulsar, as well as a large set of simulated pulsars. Our codes are publicly available for use.
\end{abstract}

\keywords{X-ray pulsar, Period estimation -- error analysis -- $Z_n^2$ and $H$ statistics}

}]


\doinum{12.3456/s78910-011-012-3}
\artcitid{\#\#\#\#}
\volnum{000}
\Year{0000}
\pgrange{1--}
\setcounter{page}{1}
\lp{1}

\section{Introduction}

Rapidly spinning pulsars constitute an active research area of astrophysics. The rotational parameters of pulsars have occupied a central role in the study of various other phenomena involving the internal and the external regions of a pulsar. High energy pulsar data is discrete, sparse and Poisson limited \citep{FermiLATcatalog}. This is a challenge to the methods which aim to measure pulsar periods. Yet over the years, period determination methods have evolved to be more efficient \citep{2022arXiv220907954B} and robust such as, Epoch-folding \citep{1990MNRAS.244...93D}, $Z_n^2$ method \citep{Buccheri1983} and $\rm{H}$-test \citep{deJager1989}. Error in frequency in time series data with white noise has been extensively discussed in many literature \citep{ransom,hare,chang,Bretthorst1988}. Despite decades of research, the field of pulsar studies in high-energy astrophysics has no commonly accepted, mathematically backed methods for error analysis, associated with period searches in Poisson limited data. Some common methods used in literature are not well scrutinised. Therefore, it is hard to associate a confidence  interval around the period estimate.

We investigated the errors in  periods recovered from pulsar data and formulated a straightforward method to reliably obtain errors. In comparison, we have shown that several previous error estimations differ by order(s) of magnitude. We performed several Monte-Carlo simulations based on ${Z_n^2}$ period estimation method, a well-established and studied method used in $H$-Test, to establish an empirical  relation between error and known parameters. An independent thorough mathematical analysis of the results was also conducted.%

The manuscript is organised as follows.
We describe the explanation of the pulsar data and period/ frequency extraction method used in Monte-Carlo simulations in Sec.~\ref{sectionmontecarlo}. It is followed by a mathematical analysis of the error in the recovered frequency in   Sec.~\ref{sec_analytical}, while Sec.~\ref{sec_powerlaw} describes the empirical power-law estimation of the error. Sec.~\ref{sec_robustresults} establishes the robustness of our formulation. This is followed by a brief overview of existing methods in Sec.~\ref{sec_comparewithothers}, the probable application and impact of our work in Sec.~\ref{sec_impact}. We conclude with a summary of our work in Sec.~\ref{sec:conclusion}.

\begin{figure}
    \centering
\begin{tikzpicture}[text width=6cm, node distance=1.7cm]
\node (start) [startstop] {Simulate time of arrivals of photons for the signal $y(t) = a + b\sin(2\pi f_0 t + \Phi)$};
\node (in1) [process, below of=start, text width=5cm, node distance=2.0cm] {Compute $Z_1^2(f)$ for an array of frequencies around the signal frequency.};
\node (in2) [process, below of=in1] { Extract $f_0$ by curve fitting a sinc-squared curve.
};
\node (pro1) [process, below of=in2] {Phase fold with recovered $f_0$  to get pulse profile and extract rest of the parameters.
};
\node (pro2) [process, below of=pro1, node distance=2.0cm] {Repeat process to obtain the distribution of all recovered parameters and measure the errors. 
};
\node (pro3) [process, below of=pro2, text width=5.5cm, node distance=2.2cm] {Implement this method across a range of source parameters to examine a relation between these parameters and the resultant errors.};
\draw [arrow] (start) -- (in1);
\draw [arrow] (in1) -- (in2);
\draw [arrow] (in2) -- (pro1);
\draw [arrow] (pro1) -- (pro2);
\draw [arrow] (pro2) -- (pro3);
\end{tikzpicture}

\caption{Flow-chart of the basic methodology to determine the uncertainty in recovered period as a function of the source parameters.\label{flow-1}}    
\end{figure}
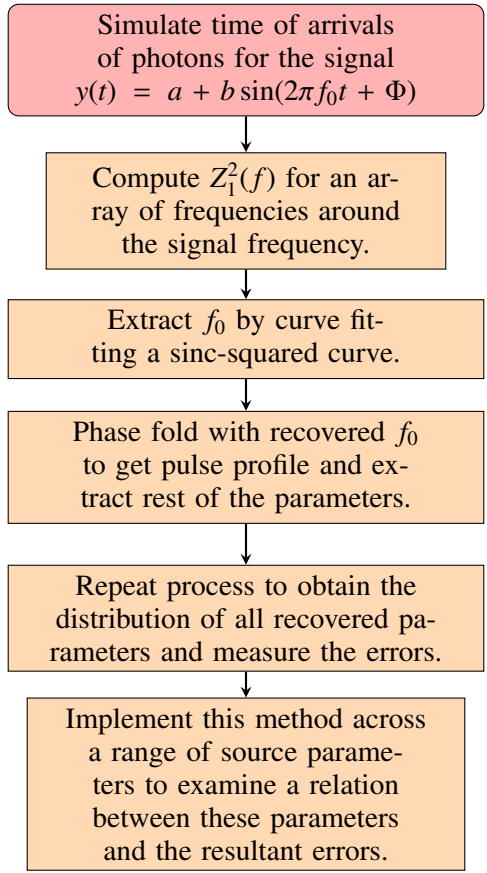

\section{Data type and simulations}\label{sectionmontecarlo}

To investigate the  uncertainty measures in recovered periods in X-ray regime, we run Monte-Carlo simulations as described in the flowchart in Fig.~\ref{flow-1}, where, Poisson--limited pulsar--like signals were simulated and their parameters were recovered using commonly used 
$Z_n^2$ method \citep{Buccheri1983}.  Multiple such realisations can give us a measure of uncertainty associated with this process.

\subsection{Pulsar Light Curves\label{lightcurve}}
Signals from high-energy pulsars are characteristic of being periodic and often very sparse. Therefore, the times of arrivals (ToA) of those high energy photons are governed by a non-stationary (and periodic) Poisson process. The ToA from a pulsar corresponding to a spin frequency $f_0$ can be modeled to follow sinusoidal Poisson process defined as, 
\begin{align}
    y(t) = a + b\sin{(2\pi f_0 t + \Phi)}. \label{sinecurve}
\end{align}
Here, $y(t)$ is the Poisson arrival rate of  photons from the  pulsar (hereafter to be referred as the light curve),  $\Phi$ is the initial phase of the light curve and $a, b$ [photon counts per sec] are light curve parameters. Furthermore, we define a measure to determine the strength of the signal which is observed for $T$ [s]. Assuming $T$ is large enough, the average number of photons received will be $aT$ with standard deviation of $\sqrt{aT}$ while the average number of photons received from the periodic component of the signal (from minimum to maximum signal) will be $bT$. We define the ratio of number of photons in the signal ($bT$) to the standard deviation in the data ($\sqrt{aT}$) as the signal strength $SS=bT/\sqrt{aT}$.

\subsection{Simulating ToA}\label{ToAsimulate}

\begin{figure}[]
\centering
  \subcaptionbox{Binned ToA which were simulated for a sinusoidal light curve $100.3 +50.1\sin( \pi t+\pi/2)$. The binning time resolution is 0.2~s. The signal  parameters here have been purposely chosen so that the variability is visually obvious. Even for such a strong signal, it can be seen that the Poisson-limited behavior of the data results in a slight deviation from the sinusoidal. \label{fig:bin}}{\includegraphics[width=0.4\textwidth]{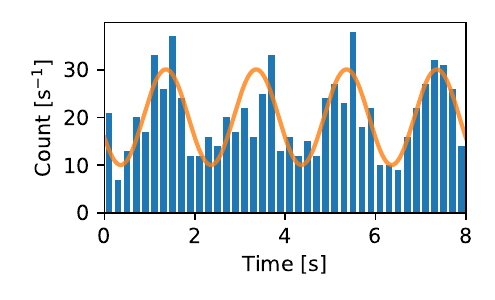}}\\
  \subcaptionbox{$Z_{1,\rm{bi}}^2$ computed around the frequency of same simulated data shown above for observation time of 100~s, using Eq.~(\ref{zn2binned}). The plot follows sinc-squared behaviour as discussed in Sec.~\ref{sec:sinc}.  \label{fig:zn2}}{\includegraphics[width=0.4\textwidth]{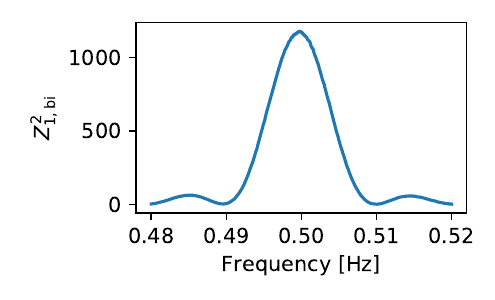}}\\
  \subcaptionbox{Pulse profile folded as per period calculated above, scaled appropriately and fit with the known light curve shape to recover the signal parameters and the initial phase. The x-axis is the phase from 0 to $2\pi$ and the y-axis is the histogram count of the number of photons within the phase resolution. \label{fig:ppf}}{\includegraphics[width=0.4\textwidth]{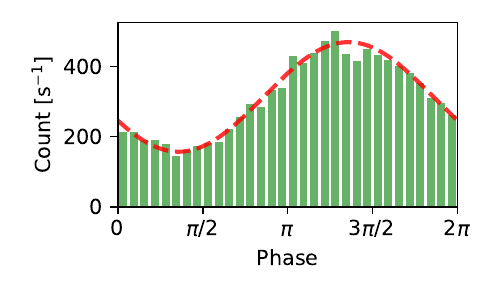}}
 \caption{Simulation process shown for a) Generating ToA (and binned for the plot), b) determination of the spin frequency using $Z_{1,\rm{bi}}^2$ method, and c) rest of the source parameters recovered using folded light curve. The simulated pulsar is chosen to be bright enough to see individual peaks.\label{sim}
 }
\end{figure}

Given the light curve and the observation time $T$, 
Python based module \texttt{Stingray}\footnote{\url{https://docs.stingray.science/}} (\citep{stingray})  has been used to simulate the ToA $(t_k)$. Here, $k$ varies from 1 to the total number of photons ($N$). The ToA obey a Poisson distribution such that the probability distribution of number of photons received between time $t_0$ and $t_0+\Delta t$, for small $\Delta t$ is given by Poiss$(y(t_0)\,\Delta t)$ distribution\footnote{Poiss($\lambda$) is a Poisson distribution with the rate $\lambda$.}. Here $y(t_0)\:\Delta t$ is the instantaneous rate of the Poisson process for that time resolution $\Delta t$. When ToA are binned they roughly follow the light curve $y(t)$ as shown in Fig.~\ref{fig:bin}. 

\subsection{Frequency extraction using \ensuremath{Z_n^2} Statistics }\label{period_using_zn2}

Given the ToA, a statistical measure of test of periodicity $Z_n^2$ can be computed as a function of frequency ($f$) as explained in \citet{Buccheri1983} using,

\begin{align}
    Z_n^2(f) = \frac{2}{N} \sum_{k=1}^{n} \left[{\left(\sum_{m=1}^{N}{\sin{(k\phi_m)}}\right)^2 + \left(\sum_{m=1}^{N}{\cos{(k\phi_m)}}\right)^2}\right]\label{zn2}.
\end{align}
Here, $n$ is the number of harmonics and $\phi_m$ is the phase corresponding to the photon with arrival time $t_m$, which is computed for a monochromatic signal as,
 \begin{align}
     \phi_m= \bm{\{}ft_m\bm{\}}2\pi,\label{phi_toa}
 \end{align}
 where $\bm{\{}x\bm{\}}$ is the fractional part of $x$. It is to be noted that Eq.~(\ref{phi_toa}) holds only when the frequency of the signal is constant. For a more general case, more corrective terms will be needed.

 Calculating $Z_n^2$ using Eq.~\ref{zn2} can be computationally expensive, as explained in \citet{stingray}, which gives the alternate expression for $Z_n^2$ by binning the phases as, 
 \begin{align}
 Z_{n,\rm{bi}}^2(f) = \frac{2}{N} \sum_{k=1}^{n} \left[\left(\sum_{m=1}^{n_b}{w_m\sin{(k\phi^{\prime}_m)}}\right)^2 + \right. \nonumber\\ \left.\left(\sum_{m=1}^{n_b}{w_m\cos{(k\phi^{\prime}_m)}}\right)^2\right]. \label{zn2binned}
\end{align}
 where $n_b$ are the number of bins the photon phases $\in [0,2\pi)$ are divided into, $w_m$ represents the number of photons in the $m^{\rm th}$ bin and $\phi'_m = 2\pi (m-1)/n_b$. 

Since the photon phases are a function of the frequency ($f$) chosen by the user, so are the $w_m$. It is to be noted that \citet{Bachetti} recommend $n_b \geq 10~n$. We chose $n_b=32$ which is the default value used in \texttt{Stingray}, which meets this requirements in our work. In the rest of this manuscript, $Z_n^2$ will be calculated by using $Z_{n,32}^2$, (\textit{i.e.,} $Z_n^2 \equiv Z_{n,32}^2$) unless otherwise specified. Furthermore, choice for the time resolution ($\Delta t$) used to simulate the ToA in Sec.~\ref{ToAsimulate} must ensure $n_b ~\Delta t \le 2\pi/f_0$ to avoid aliasing effects.

 Since $Z_n^2$ statistics peaks around the signal frequency, the   frequency of the signal can be recovered by computing the statistic on an array of equidistant  frequencies ($f_k$) \citep{stingray} with a range {$1/T$}. We pick a time resolution  {$1/(250~T)$}, to ensure good sampling in that region. The array is centered around the injected frequency. We add a random offset to the injected frequency to ensure that it is not contained on the sampling array, to avoid any biases it may cause. For computational simplicity, we compute $Z^2_1$ 
 statistics, a special case of Eq.~(\ref{zn2binned}), also known as Rayleigh test \citep{Rayleigh1919}. Fig.~\ref{fig:zn2} shows plot of $Z_{1}$ as a function of frequency for a simulated light curve as shown in Fig.~\ref{fig:bin}.

\subsection{Extracting spin frequency and other parameters\label{sec:sinc}}

Based on the $Z_{1}^2$ statistics, the frequency at which $Z_{1}^2(f)$ achieves its maximum value, can be used to extract the frequency of the signal. This method however is limited to resolution of frequency at which $Z_{1}^2$ statistics is computed. It is observed that the statistic follows a sinc-squared distribution around the signal frequency (Fig.~\ref{zn2overlayplot}), giving a computationally effective method to recover the frequency \citep{Leahy1983b,Leahy1987,stingray} 
A generic sinc-squared function can be described as,
\begin{align}
    Z(f) = A \cdot \rm{sinc}^2\left(\frac{f-f^{\prime}}{W}\right)\label{y_sinc},
\end{align}
where the parameters $A,W$ are the amplitude and width of the function, $f^{\prime}$ is the position where the function takes its maximum value, sometimes referred to as the ``mean'' of the sinc-squared function. The best fit for $f^{\prime}$ gives us the recovered frequency (hereafter, the recovered parameter will be referred with subscript `r'). Based on the recovered $f_r$ the phases of each ToA can be computed. Binning these phases allows us to recover the remaining light curve parameters $(a,b, \Phi)$ from the phase-folded lightcurve as shown in Fig.~\ref{fig:ppf}. 

Hereafter, we will refer to the best-fit ``recovered'' light curve parameters as $(a_r,b_r,\Phi_r)$. Further details on the benefits of curve fitting at the peak of detection statistics are elaborated in Sec.~\ref{max-cur}.

\subsection{Generating multiple realisations}\label{multiple_simulations}

\begin{figure}
\centering
  \includegraphics[width=0.4\textwidth]{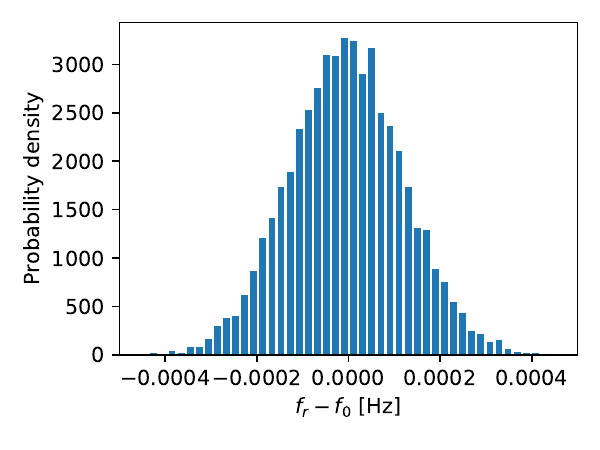}
  \caption{Probability density of deviation in recovered frequency from the source frequency $(f_r - f_0)$ using sinusoidal curve for $(a,b,f_0,T)=(10, 7, 1, 200)$. 
  The distribution from $n_{\rm{sim}}=10\,000$ realisations, follows a Gaussian distribution. }
  \label{distribution_omegas}
\end{figure}

Using the described methodology, we can simulate pulsar-like data for a given set of source parameters $(a,b,f_0,\Phi)$ and observation time $T$, and recover the parameters $(a_r,b_r,f_r,\Phi_r)$ from the simulations. By repeating these simulations for $n_{\rm{sim}}=500$ times, we obtain multiple realizations of the recovered parameters $(\Lambda_{r,k})$ for the $k^{\rm th}$ simulation, where $\Lambda \in {a,b,f_0,\Phi}$ and $k$ takes integer values from 1 to $n_{\rm{sim}}$. To avoid any biases, the phase ($\Phi$) was randomly sampled from a uniform distribution in every simulation. For a sufficiently large value of $n_{\rm{sim}}$, a histogram of $\Lambda_r$ as shown in Fig.~\ref{distribution_omegas} can be considered as the true distribution of the recovered parameter and can thereby be used to estimate its variance. 

\begin{figure}
    \centering
    \includegraphics[width=0.4\textwidth]{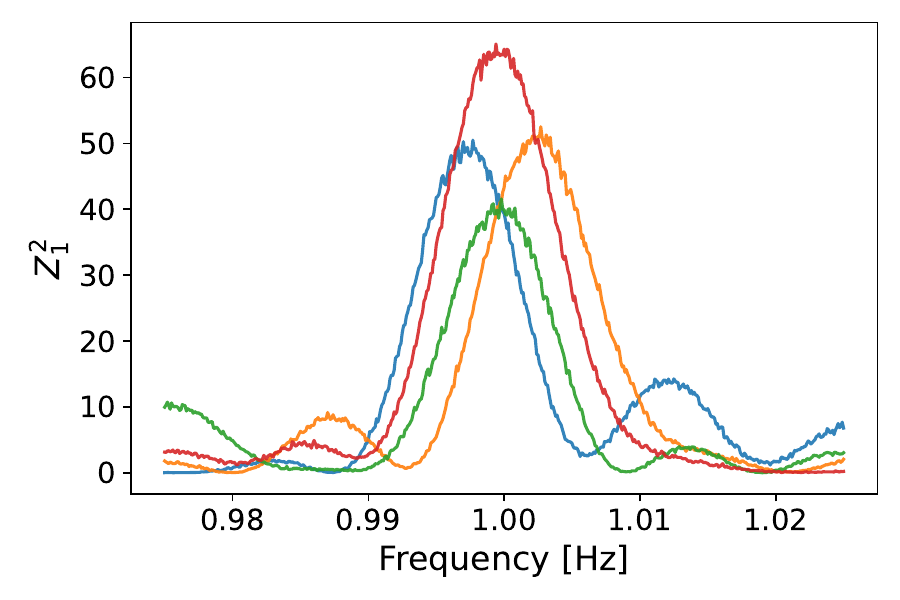}
    \caption{$Z_1^2(f)$ plotted for the multiple realisations of the same source parameters signal of $(a,b,f_0,T) = (50,5,1,200)$. The plot shows variation in the peak position and height for different realisations. All the realisations of $Z_1^2(f)$ roughly follow to sinc-squared variation with frequency. If we fit this with a sinc-squared function, the resultant uncertainty in recovered central frequency for each of these plots is significantly smaller than the change of the peak position itself across different simulations.
    } 
    \label{zn2overlayplot}
\end{figure}

The Fig.~\ref{zn2overlayplot} depicts $Z_1^2$ from several realizations, providing insight into the variability in the estimated value of $f_r$. It is seen that there can be significant shifts in the entire $Z_1^2$ distribution across different realisations of the same underlying process.

Consequently, the uncertainty in estimating $f_r$ from fitting the sinc-squared function to such a distribution does not accurately reflect the actual uncertainty in measurement of $f_r$.

In Section~\ref{sec_analytical}, we present a justification for the shift in the peaks observed in $Z_1^2$ statistics, attributing it to the contribution of Poisson noise in the signal.

We define the error in measuring $\Lambda_r$ as the median of absolute deviation ($\rm{MAD}$) from the source parameter $\Lambda_0$, as given by,

\begin{align}
    \rm{MAD}(\Lambda_r) = \underset{1 \leq k \leq n_{\rm{sim}}}{\text{Median}}\left(\left|\Lambda_{r,k} -\Lambda_0\right|\right)\label{MADdef}.
\end{align}

The selection of this error measure can be justified based on its robustness against statistical outliers and its dependence on the deviation from the injected parameters rather than from the mean of the recovered parameters. We assume that the probability density of $f_r$ follows a Gaussian distribution with a mean of $f_0$ and a standard deviation of $\sigma(f_r)$, as supported by the distribution shown in Fig.~\ref{distribution_omegas}. For such a Gaussian distribution, a region of approximately $\sigma(f_r)/\sqrt{2}$ from the mean will have a 50 per cent confidence interval\footnote{Note that $\sigma/\sqrt{2}$ has a confidence interval of 52.1 per cent, which is a reasonable approximation.}. This is identical to the confidence interval obtained using MAD$(f_r)$.

Hence, we can approximate the following relation,
\begin{align}
    \sigma(f_r) \approx \sqrt{2}~\rm{MAD}(f_r).\label{sig_mad}
\end{align}
In the next section, we analytically investigate the behaviour of these uncertainties as a function of source parameters and the observation time, based on which will run Monte-Carlo simulations and test the results.

\section{Analytical Calculation}\label{sec_analytical}
 
As discussed, it seems that the major contribution of this uncertainty comes from the lateral variation on sinc-squared parameters due to the stochastic nature of the variables $w_n$, as defined in Eq.~(\ref{zn2binned}). We therefore, analytically study the propagation of Poisson noise in measuring $f_r$ using the $Z_n^2$ method, when the number of cycles observed is sufficiently high.

For simplicity, we start with a sinusoidal light curve and hence, go up to $n=1$ harmonics; we will later generalise for higher harmonics. It is observed that Eq.~\ref{zn2} can be alternately written as
\begin{align}
    Z_n^2(f) = \frac{2}{N} \sum_{k=1}^{n} \left|\:\sum_{m=1}^{N}{e^{-ik\phi_m}}\right|^2; \text{where, } i=\sqrt{-1}.\label{newzn2}
\end{align}
Considering a deterministic pulsar light curve $g(t)$, which implies no Poisson noise and observation time $T$, the number of photon arrivals in a small interval $t$ to $t+\Delta t$ is exactly $g(t)~\Delta t$. Using Eq.~(\ref{phi_toa}), the phase $[\phi(t)]$ corresponding to these photons for a frequency ($f$) is
$2\pi\:\{f t\}$. Hence, Eq.~\ref{newzn2} can be rewritten as
\begin{align}
    Z_n^2(f) = \frac{2}{N} \sum_{k=1}^{n} \left|\:\sum_{m=1}^{M}{w(t_{m,\rm{bin}})e^{-ik\phi^{\prime}(t_{m,\rm{bin}})}}\right|^2,
\end{align}
where $t_{m,\rm{bin}} = (m-1)\Delta t$, $M=T/\Delta t$ and $w(t_{m,\rm{bin}})$ is the number of photon arrivals in $[t_m, t_m + \Delta t)$. When $\Delta t$ is arbitrarily small, we can now rewrite the above equation as,
\begin{align}
    Z_n^2(f) =  \frac{2}{N} \sum_{k=1}^{n} \left|\:\int_{0}^{T}{g(t)e^{-2\pi ikf t}dt} \right|^2. \label{zn2int}
\end{align}
Since $g(t)$ is a real function, we can write,
\begin{align}
Z_1^2(f) = (2/N)|G(f)|^2,\label{FT}
\end{align}
where $G(f)$ is the Fourier transform of $g_{\rm rec}(t)$ given by 
\begin{align}
    g_{\rm rec}(t) = g(t) \cdot \rm{rect}\left(\frac{t - T/2}{T}\right).\label{g_rt}
\end{align}
Here rect($x$) is a top-hat window function  
which takes the value 1 in the window $x \in [-1/2,1/2]$ and is 0 elsewhere. When $g(t)$ is same as $y(t)$ given by Eq.~\ref{sinecurve} and the number of observed cycles is large, $Z_1^2(f)$ can be written as

\begin{align}
Z_1^2(f) &\approx \frac{Tb^2}{2a}\rm{sinc}^2\left[T\pi(f-f_0)\right] + \frac{T(2a)^2}{2a}\rm{sinc}^2\left(\pi Tf\right)  \nonumber\\ &\quad+ \frac{Tb^2}{2a}\rm{sinc}^2\left[T\pi(f+f_0)\right].\label{zn2sine}
\end{align}
It is worth noting that we neglect one term in this representation a more comprehensive discussion on this term can be found in Sec.~\ref{sec:fullformula}. Around $f=f_0$, the second and third terms diminish.  and hence we can simplify the above equation as,  
\begin{align}
  Z_1^2(f) &= \frac{Tb^2}{2a}\rm{sinc}^2\left[T\pi(f-f_0)\right]\label{sinc-zn2},\\
  &= A~\rm{sinc}^2\left(\frac{f-f_0}{W}\right). \label{zsin2}
\end{align}

Here, the amplitude ($A$) and width ($W$) of the sinc-squared as defined in Eq.~(\ref{y_sinc}) are $Tb^2/2a$ and $1/(T\pi)$ respectively. 
Taylor expanding Eq.~\ref{zsin2} around $f_0$
\begin{align}
    A~\rm{sinc}^2\left(\frac{f-f_0}{W}\right) 
    &= A - \frac{A}{3W^2}(f-f_0)^2 \nonumber\\ &\quad+ \mathcal{O}\left[A\left(\frac{f-f_0}{W}\right)^4\right],\label{AW2}\\
    &= \frac{T b^2}{2a} - \frac{b^2 \pi^2 T^3}{6 a}(f-f_0)^2 \nonumber\\ &\quad+ \mathcal{O}\left[T^5\left(f-f_0\right)^4\right].\label{abt2}
\end{align}

It can also be noted that higher order terms in above equation can be ignored when $\left|f-f_0\right| \lesssim W$. Therefore the curve fitting routine could be fine-tuned to be applied for region $\left|f-f_0\right| = W$. Continuing on the same line of reasoning for a generic pulse profile we can define the light curve as a Fourier series defined as,
\begin{align}
g(t) = a_0 + \sum_{i=1}^{n} b_k \sin(2\pi kf_0 t + \Phi_k),\label{fourier}
\end{align}
Eq~(\ref{zn2int}) can be expressed as,
\begin{align}
    Z_n^2(f) = \sum_{k=1}^{n}\frac{Tb_k^2}{2a}\rm{sinc}^2\left[{T\pi k}(f-f_0)\right]\label{sinczn2},
\end{align}
and Taylor expanding it gives us an analytical expression for any generic light curve as,
\begin{align}
    Z_n^2(f) &= \cfrac{T\sum_{k=1}^{n}b_k^2}{2a} - \cfrac{T^3\pi^2\sum_{k=1}^{n}k^2b_k^2}{6 a}\left(f-f_0\right)^2 \nonumber\\ &\quad+ \mathcal{O}\left[T^5\left(f-f_0\right)^4\right].\label{zn_taylor}
\end{align}
Although these expressions are derived for a deterministic pulsar data however, even with the introduction to Poisson governed ToA, it is trivial to show that the expectation of $Z_n^2(f)$ is same as described in Eq.~(\ref{zn_taylor}) as detailed in Sec.\ref{subsec:expectation_variance}.  
We now investigate how the Poisson noise affect the recovered frequency.

\subsection{Variation in   frequency}\label{varomega}
To investigate the variance in the recovered frequency due to the spread of the mean of the  sinc-squared curve used to fit the $Z_n^2(f)$ from a simple sinusoidal signal, it is reasonable to assume that the mean is where $Z_n^2(f)$ peaks. Thus, $h(f_r)=0$ where,
\begin{align}
    h(f^{\prime}) = \left.\frac{d}{df}Z_n^2(f)\right|_{f=f^{\prime}},
\end{align}
and $f_r$ is the recovered  frequency. When the ToA is free of Poisson noise, $h(f_0)=0$ and $f_r=f_0$. On the introduction of Poisson noise in data, $f_r$ is no longer constrained to be the same as $f_0$ and may be written as $f_r=f_0+\Delta f$. For simplification, we employ time-domain analysis and assume that the ToA are binned with a time resolution $\Delta t$ such that the total observation time $T$ is an integral multiple of $\Delta t$. This results in an array $\{g_m\}$, representing the number of photons received in the $m^{\rm{th}}$ time bin, with $m$ ranging from 1 to $N_d$ and $N_d = T/\Delta t$.
 For no Poisson noise and infinitesimally small $\Delta t$, $g_m\approx g_0(t_m)\Delta t$ where $g_0(t)$ is the true periodic light curve of the pulsar and $t_m=(m-1)\Delta t$. However, due to the presence of Poisson noise, we may write $g_m=g_{m,0}+\Delta g_m$ where $g_{m,0}= g_0(t_m)\Delta t$. Also, it is known that in such a case \begin{align}
    Z_n^2(f) = \frac{2}{N} \sum_{k=1}^{n} \left|\:\sum_{m=1}^{N_d}{g_m e^{-ik\phi(t_{m,\rm{bin}})}}\right|^2,
\end{align}
where $N$ is the total number of photons received and $\phi(t_{m,\rm{bin}})$ is the same as in Eq.12. Thus, $Z_n^2(f)$, $h(f)$ and $f_r$ depend on $\{g_m\}_{m=1}^{N_d}$. Let us define
\begin{align}
     h(f,\{g_m\}) = \frac{\partial}{\partial f}Z_n^2(f, \{g_m\}).
\end{align}
We know that $h(f_0, \{g_{m,0}\})=0$ and $h(f_r, \{g_{m}\})=0$. Assuming $\Delta f$ and $\Delta g_m$ to be reasonably small and carrying a first-order Taylor expansion around $(f_0,\{g_{m,0}\})$, it can be seen that
\begin{align}
\frac{\partial h}{\partial f}\cdot \Delta f + \sum_{m=1}^{N_d}{\Delta 
 g_m\cdot \frac{\partial h}{\partial g_m}}=0,  
\end{align}

where the derivatives are evaluated at $f=f_0$ and $\{g_m\} = \{g_{m,0}\}$. Thus, 
\begin{align}
    {\rm{Var}}(\Delta f) = \left(\frac{\partial h}{\partial f}\right)^{-2} \sum_{m=1}^{N_d}{{\rm{Var}}(\Delta g_m)\cdot \left(\frac{\partial h}{\partial g_m}\right)^2},\label{eq:var_dirty}
\end{align}
since $\{g_m\}$ are independent random variables. Variance in $\Delta g_m$ is the same as the variance in $g_m$, which is equal to $g_{m,0}$ owing to the Poisson nature of the noise. So, 
\begin{align}
    {\rm{Var}}(\Delta f) = \left(\frac{\partial h}{\partial f}\right)^{-2} \sum_{m=1}^{N_d}{g_{m,0}\cdot \left(\frac{\partial h}{\partial g_m}\right)^2},\label{analytical_sum_discrete}
\end{align}
We know that when $g_m = g_{m,0}$ or equivalently there is no Poisson noise, then $Z_n^2(f)$ can be written as $A_0\cdot{\rm{sinc^2}}((f-f_0)/W_0)$. Thus, $\partial h/\partial f$ in the equation above can be substituted by $-A_0/3 W_0^2$ and hence, it is trivial to show that ,
\begin{align}
    {\rm{Var}}(\Delta f) = \left(\frac{A_0}{3W_0^2}\right)^{-2} {\rm{Var}}[h(f_0)].
\end{align}

It should be noted that when the right hand side of Eq.~(\ref{analytical_sum_discrete}) is calculated for arbitrarily large $N_d$, the result can be rewritten in terms $A_0$ and $W_0$. Then we can write,
\begin{align}
    {\rm{Var}}(f_r) = \sigma^2(f_r) = \frac{3W_0^2} {{A_0}} =\frac{6a}{b^2\pi^2T^3}\label{AWpowerlaw}.
\end{align}
We notice a similarity between the Var($f_r$) in Eq.~(\ref{AWpowerlaw}) and the second coefficients in Eqs~(\ref{AW2}) and~(\ref{abt2}). This relationship parallels the findings of \citep{ransom}, where a similar association between frequency error and the second derivative was observed in time series influenced by white noise. By substituting $\rm{Var}(\Delta g_m) = \sigma^2$ for white noise in Eq.~\ref{eq:var_dirty}, where $\sigma^2$ represents a constant variance typical of white noise approximated as $a$, it can be demonstrated that the error in the recovered frequency for large $T$ converges to the expression in Eq.~\ref{AWpowerlaw}.

Here $\sigma(f_r)$ is the standard deviation (error with $\sim68$ percent confidence) in $f_r$. Since, MAD results in error with 50 percent confidence, we use the relation mentioned in Eq.~(\ref{sig_mad}) to approximate,
\begin{align}
\text{MAD}(f_r) &\approx \frac{\sigma(f_r)    }{\sqrt{2}} = \frac{\sqrt{3}}{\sqrt{2}}\frac{W_0}{\sqrt{A_0}} = \frac{\sqrt{3a}}{\pi b T^{3/2}}. \label{madpl}
\end{align}

Continuing this formulation for generic light curve as in Eq.~(\ref{fourier}),  we use the fact that for $k>1$),
\begin{align}
    Z^2_k(f) - Z^2_{k-1}(f) &= \frac{Tb_k^2}{2a}\rm{sinc}^2\left[Tk\pi (f-f_0)\right].\label{harm}
\end{align}
Therefore, we can safely comment on the variance in estimating $f_{r,k}$ for each $k>1$ in above equation will be given by, 
\begin{align}
    \sigma_k^{2}(f_r) = \frac{6 a}{b_k^2 k^2 \pi^2 T^3}.
\end{align}
Taking the variance of the arithmetic mean of $f_{r,k}$,  which is given by,
\begin{align}
\sigma^2(f_r) = {\frac{ 1 }{\sum_{i=k}^n \sigma_k^{-2}}},
\end{align}
results in the variance being the inverse of the second coefficient in Eq.~(\ref{zn_taylor}) as,
\begin{align}
    \sigma^2(f_r) = \cfrac{6 a}{T^3\pi^2\sum_{k=1}^{n}k^2b_k^2}.\label{gen_sigma}
\end{align}
This further justifies our previous hypothesis, as the variance is again similar to the second term in Eq.~(\ref{zn_taylor}). Furthermore, \citet{1996A&AS..117..197L} explains that Eq.~(\ref{gen_sigma}) should be a direct generalisation of Eq.~(\ref{AWpowerlaw}) as ``\textit{higher harmonics are folded with
k cycles over one fundamental pulse cycle}''.
Hence, we conclude that the shape of a small region around the peak of the $Z_n^2$ can give us the estimation of the error in frequency. Fitting the $Z_n^2$ distribution with a sinc-squared function (or Gaussian or quadratic) gives the information for the error. We preferred sinc-squared function as it has shown consistency in our simulations. We therefore present a mathematical formulation of the light curve shape independent MAD as,
\begin{align}
    \text{MAD} = \frac{\sqrt{3}}{\sqrt{2}}\frac{W_r}{\sqrt{A_r}},
\end{align}
which we will refer to as a $AW$ power-law. We further numerically test this formulation on simulated and observed pulsar data.

\subsection{H-Test}
So far we have treated number of harmonics as a user choice but, the $H$-test as described in \citet{deJager1989} and \citet{hart1985choice} eliminates this degree of freedom as,
\begin{align}
    H &\equiv \underset{1 \leq m \leq 20}{\text{Max}}\left(Z_m^2 - 4m +4\right), \label{htest}\\
    &= Z_M^2 - 4M +4,
\end{align}
for some $M$ between 1 and 20. Using the obtained statistics a probability density function can be associated with it as discussed in \citet{2010A&A...517L...9D} as,
\begin{align}
    \text{Prob}(H>q) = \exp{(-0.4~q)}.
\end{align}
Using the analytical expression for the $Z_n^2$ for a known light curve of a pulsar, we can express the $H$-statistics as,
\begin{align}
    H &\equiv \cfrac{T\sum_{k=1}^{M}b_k^2}{2a} - \cfrac{T^3\pi^2\sum_{k=1}^{M}k^2b_k^2}{6 a}\left(f-f_0\right)^2 -4M +4,
\end{align}
and hence, extend it to a corresponding probability density function. In this study however, we will still treat number of harmonics as a user dependent choice.

\section{Power Law estimate}\label{sec_powerlaw}

\begin{table}
	\centering
	\caption{Grid of source parameters which were regularly sampled within the range of column~2 and~3. Column~4 mentions if the sampling was in linear or log scale along with number of elements sampled in column~5. A large dynamic parameter space was used to avoid any selection biases (with the condition $a>b$).\label{tab:grid_composition}}
	\begin{tabular}{lp{1cm}p{1cm}cr} 
		\hline
		Parameter & Lower Bound & Upper Bound & Sampling & Number\\
		\hline
		$f_0~[Hz]$ & 1 & 10 & Lin & 1024\\
		$a~[s^{-1}]$ & 75 & 750 & Log& 1024\\
		$b~[s^{-1}]$ & 25 & 250 & Log & 1024\\
		$T~[s]$ & 1000 & 5000 & Log& 1024\\
		\hline
	\end{tabular}
\end{table}

The analytical formulations derived in Eq.~(\ref{madpl}) tells us that the error $f_r$ recovered using $Z_1^2$ due Poisson nature of the data has a simple power-law behaviour. We numerically test this hypothesis by measuring the MAD$(f_r)$ and comparing it with the formulation. We run Monte-Carlo simulations for various source parameters to independently investigate the power-law as well as the parameter space where the formulation is more reliable.  We then, uniformly sample the source parameters as described in Table~\ref{tab:grid_composition}. We define the quantity, error ratio (ER) that compares numerically obtained errors with the power-law as,
\begin{align}
    \rm{ER} &= \frac{\text{Numerically computed MAD($f_r$)}}{\sqrt{(3 a/\pi^2)}~ b^{-1}T^{-1.5}},\label{error_ratio}
\end{align}
Further, we also explore if the error in other source parameters $(a, b, \Phi)$ [when recovered using the methodology described in  Fig.~\ref{flow-1}], be expressed as a power-law.

\subsection{Simulations and Power Law Estimation}

We begin our investigation of whether the error in the recovered parameters, as described in Sec.~\ref{sec:sinc} can be approximated by a power-law as a function of the source parameters $\Lambda \in \{a,b,f_0,\Phi\}$.  
For a given parameter $\Lambda$, we define its power-law using the exponent $P_a, P_b, P_{f}$, $P_T$, and  the constant $M_0$ as,
\begin{align}
    {\rm{MAD}}(\Lambda_r) = M_0~\left(a^{P_a}\right)\left(b^{P_b}\right)\left(f_0^{P_{f}}\right)\left(T^{P_T}\right).\label{generalpowerlaw}
\end{align}

\begin{figure}
\centering
  \subcaptionbox{
  Estimating the exponent ($P_a$) of $a$ in the power-law representation of the $\text{MAD}(f_0)$, as outlined in Eq.~(\ref{generalpowerlaw}), while maintaining the parameters $(b,f_0,T)$ constant and varying only $a$. \label{fig:bpow}  }{\includegraphics[width=0.45\textwidth]{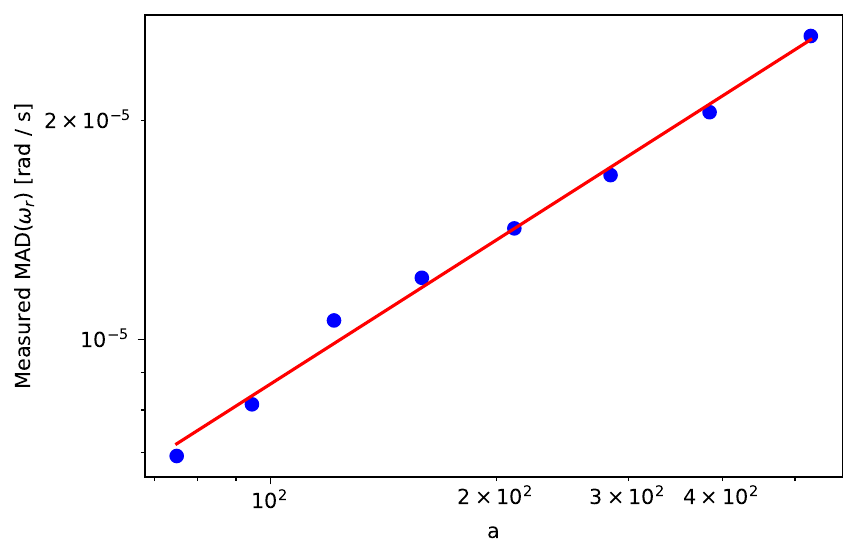}}
  \subcaptionbox{$P_a$ exponent recovered ($y$-axis) for different source parameters (simulation index on $x$-axis). The plot is colour coded with number of cycles of injected pulsar signal. The Blue line is the mean of all the exponent measured in each simulation however, clipping the low cycle data we get a more consistent average as marked by red dotted line. \label{fig:bpowss}}{\includegraphics[width=0.45\textwidth,trim={0 20 0 20},clip]{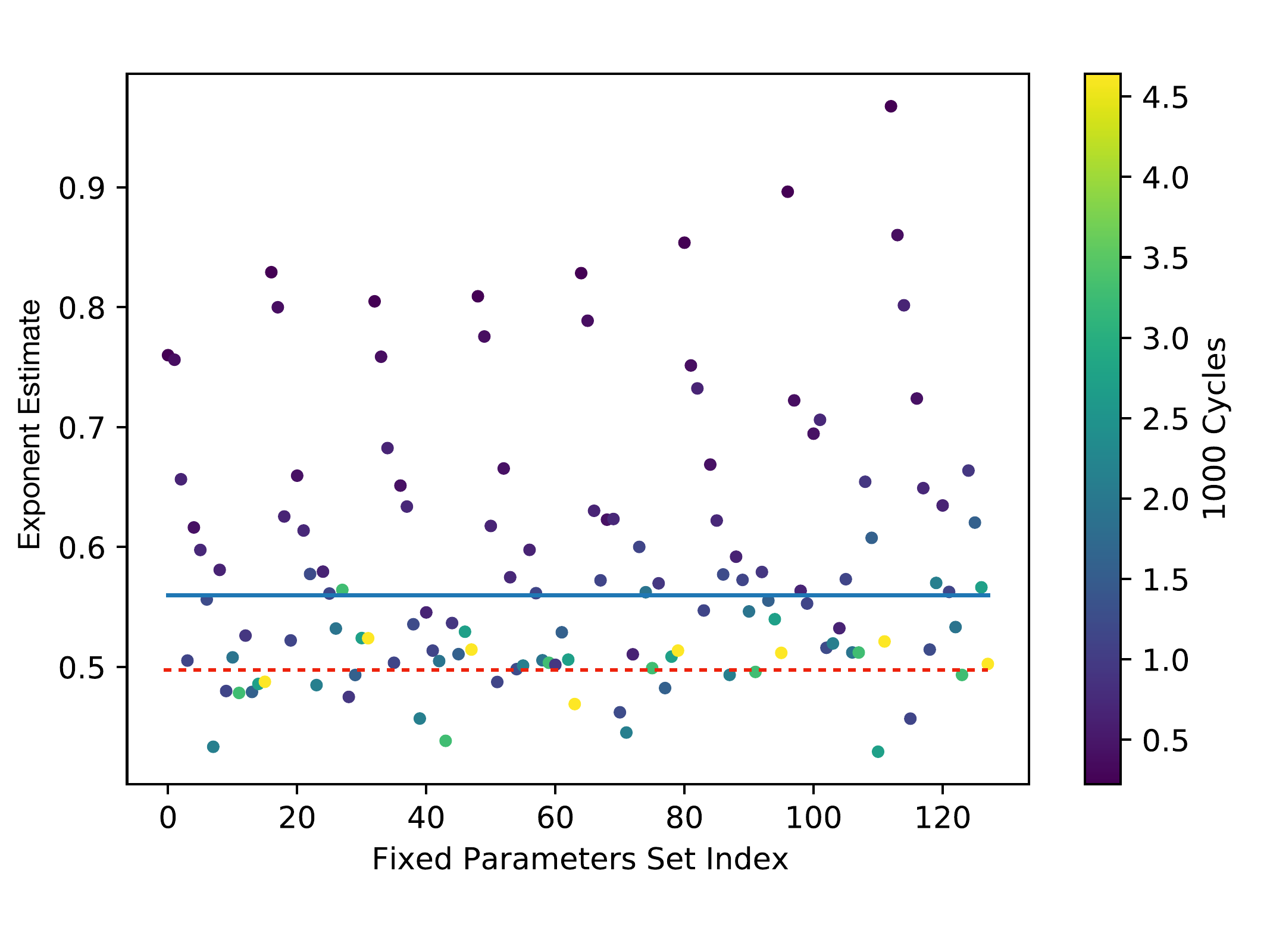}}
  \caption{Empirically estimating the power-law relationship between the parameter $a$ and the $\text{MAD}(f_0)$ for a variety of fixed $(b,f_0,T)$ sets. \textit{Top: } plot shows results of curve fitting routine for a single simulation, while the \textit{Bottom: } displays a distribution of recovered exponent for a range of parameters, to explore the dependence of this correlation on the source parameters. The findings suggest that the power-law $\text{MAD}(f_0) \propto a^{0.5}$ is consistent, given sufficient observation time.}
  \label{power-fig}
\end{figure}

It is to be noted that the power-law parameters are different for different $\Lambda$. To obtain the exponent of (say) $a$ in the power-law for ${\rm{MAD}}(f_r)$, we select sets of source parameters with the same $(b,f_0,T)$ but different $a$. The value of $P_a$ is obtained by curve fit of $y_{\rm{fit}}(a)=C_0 a^{P_a}$  to  $\rm{MAD}(f_r)$ with the errors obtained in the simulations as shown in Fig.~\ref{fig:bpow}.

This estimate fluctuates as the fixed set $[b,f_0, T]$ also varies in every simulation. Fig.~\ref{fig:bpowss} shows, different values of $P_a$ for 
different combinations of $[b,f_0, T]$.

 On closer inspection it is found that the outliers (purple data points) in  Fig.~\ref{fig:bpowss} are mostly due to the shorter number of cycles ($f_0T$) observed in the simulation. Clipping the data by removing exponents measured from simulation with less than 2000 cycles of signal, the $P_a$ becomes a more consistent value near 0.5 .  
 
\begin{figure}
\centering
  \includegraphics[width=0.45\textwidth]{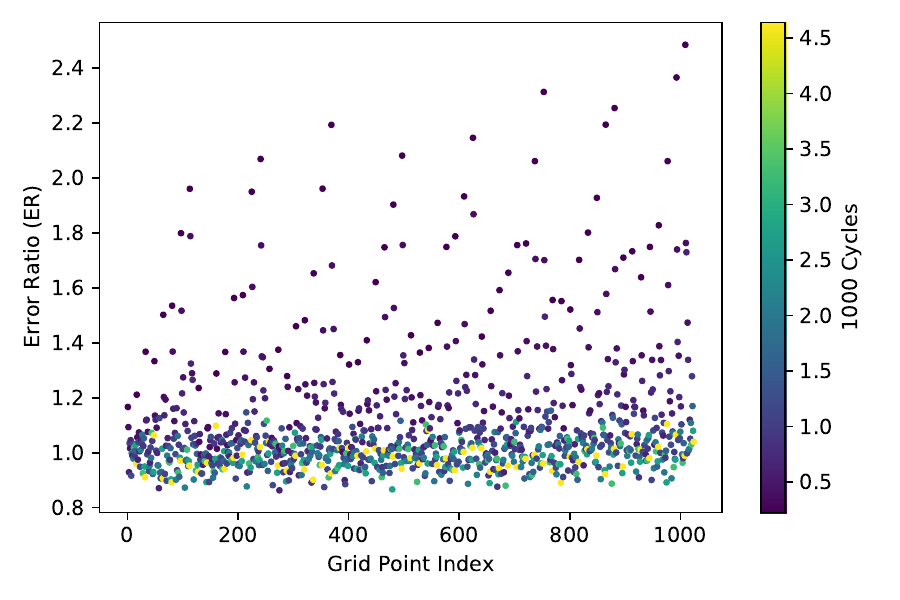}
  \caption{Plot representing the ratio of error measured from the numerical simulations and from the simplified power law as described in Eq.~(\ref{error_ratio}). It is evident that low number of cycles are common to all the outliers where the power-law underestimates uncertainty.}
  \label{ratioplot}
\end{figure}

 Following the same methodology, the rest of the exponents in the power-law can be estimated. Once all the exponents are known, computing $M_0$ is trivial as it is the only unknown left in Eq~(\ref{generalpowerlaw}). The clipped average of constants computed from all the simulation gives us the estimate of the constant and hence, the complete form of empirical power law for a given MAD($\Lambda_r$). Repeating the entire process for each $\Lambda \in \{a,b,f_0,\Phi\}$ gives us the desired power-law as shown in Table.~\ref{power}. Thus, the power-law empirically derived through simulations is within the margin of error close to the analytical formulation,
\begin{align}
    {\rm{MAD}}(f_r) = 0.55~ a^{0.5}b^{-1}T^{-1.5}.\label{simplePL}
\end{align}
However, it is reliable when the observed cycles are sufficiently large, otherwise some more correction terms may need. 
When the Eq.~(\ref{simplePL}) is tested with several source parameters, the Fig.~\ref{ratioplot} similar results, that the formulation is accurate with 10 per cent tolerance, when observed for 2000 cycles or more. It is to be further noted that the 2000 cycle threshold is not a universal criteria. As shown in a separate study it varies with the choice of source parameters but, the power-law is more reliable when observation times is sufficiently high.

\begin{table*}
\caption{\label{power} Parameters of the power-law as described in Eq.~(\ref{generalpowerlaw}). From first row it can be concluded that the empirical power-law for estimating the error in measuring   frequency using $Z_1^2$ method (with 50 per cent confidence) is $\rm{MAD}(f_r)\approx0.55~a^{0.5}b^{-1.03}f^{0.01}T^{-1.52}$}
	\begin{tabular}{lccccr} 
		\hline
		$\Lambda$ & $M_0$ & $P_a$ & $P_b$ & $P_{f}$ & $P_{T}$\\
		\hline
		$f_0$ & $0.55\pm0.1$ & $0.50\pm0.03$ &$-1.03\pm0.04$ &$0.01\pm0.02$ & $-1.52\pm0.03$\\
		\hline
	    $a$ & $0.9\pm0.3$ & $0.50\pm0.07$ &$0.01\pm0.05$ &$0.02\pm0.04$ & $-0.50\pm0.06$\\
		$b$ & $1.3\pm0.3$ & $0.39\pm0.05$ &$0.18\pm0.02$ &$-0.04\pm0.03$ & $-0.43\pm0.02$\\
		$\phi$ & $5.3\pm0.1$ & $0.12\pm0.08$ &$0.02\pm0.01$ &$-0.37\pm0.06$ & $0.00\pm0.01$\\
		\hline
	\end{tabular}
\end{table*}

\subsection{Further checks}

Before we proceed, it is important to verify that the distribution, error measure and the power law established so far, are independent of some of the choices made in Sec.~\ref{sectionmontecarlo}.

We investigated the power-law while varying the {$n_b \in \{16,32, 64, 128\}$}. We find that the power-law shows similar behaviour for all values of $n_b$ used, within margin of error. It is therefore reasonable to establish that the choice $n_b$ does not bias the result and it can be used as a good trade-off for computational resources with minor inaccuracies. 

Furthermore, we assumed that the number of simulations $n_{\rm{sim}}$ are large enough to showcase the true distribution and  therefore the true measure of uncertainty. To verify this, we investigate the stability of the measure quantity MAD as a function of $n_{\rm{sim}}$.
We observed that for the different elements in the grid, the measured error fluctuates within {10 per cent} for $n_{\rm{sim}} \gtrsim 300$. Hence, for the Monte Carlo simulations used, $n_{\rm{sim}}=500$ is justified.

Lastly and most importantly, we need to investigate and check the biases caused by choosing a simplified $y(t)$ and therefore limiting the harmonic to $n=1$ in the $Z_n^2$ method. We will further study the extension of the power-law for a generic light curve and for higher harmonics in the next section. In the next section we study the possibility of extension of the power-law for generic light curves involving higher harmonics.

\section{Robustness of the formulation}\label{sec_robustresults}
The variance of the recovered frequencies (using methodology discussed in Sec.~\ref{multiple_simulations}), can be attributed to many factors which we can describe as,

\begin{align}
    \rm{Var}(f_r) &= [\rm{Var}(f_r)]_{\rm{Poiss}} + [Var(f_r)]_{\rm{fit}} + \cdots.\label{varfr}
\end{align}
Here, $[\rm{Var}(f_r)]_{\rm{Poiss}}$ represents the variance in the shift of the peak of $Z_n^2$ due to Poisson noise in the Time of Arrival (ToA), while $[\rm{Var}(f_r)]_{\rm{fit}}$ denotes the variance in estimating the curve fitting parameters. Empirically, we find that $[\rm{Var}(f_r)]_{\rm{Poiss}}$ is nearly identical to $\rm{Var}(f_r)$, suggesting that all other contributing terms are relatively insignificant.
We have formulated an error measure for a generic light curve, and its efficacy has been tested on numerical simulations of sinusoidal light curves. We aim to further validate the robustness of this measure by testing it on different light curve shapes, real data artifacts, such as gaps in the data, and, most significantly, X-ray pulsar data from the \emph{AstroSat} satellite.

\subsection{Robustness towards different shapes\label{shapes}}

\begin{figure}
    \centering
    \includegraphics[width=0.48\textwidth]{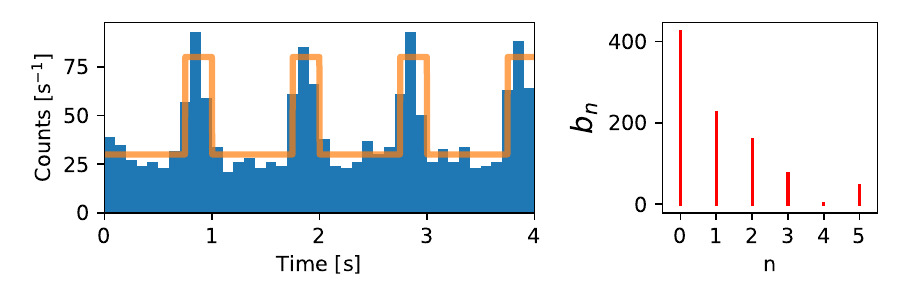}\\
    \includegraphics[width=0.48\textwidth]{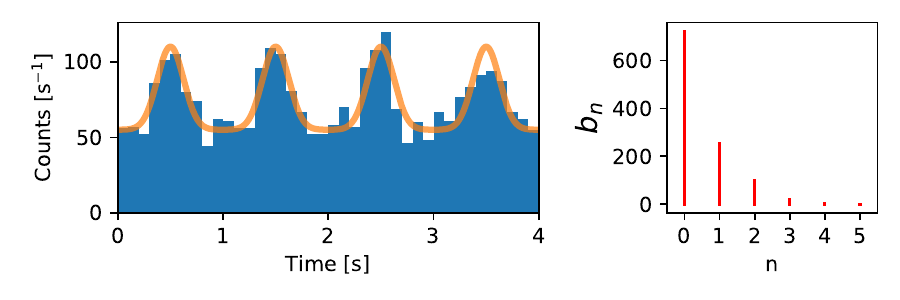}
    \caption{\textit{Top to bottom:} First column shows, binned ToA for rectangular and Gaussian shaped light curves respectively, as described in Sec.~\ref{shapes}. In the second column coefficient of Fourier series $b_n$ for different harmonics $n$, as described in Eq.~(\ref{fourier}), where $b_0 = a_0$ .}
    \label{fig:shape_toa}
\end{figure}

\begin{figure*}
    \centering
    {\includegraphics[width=0.95\textwidth]{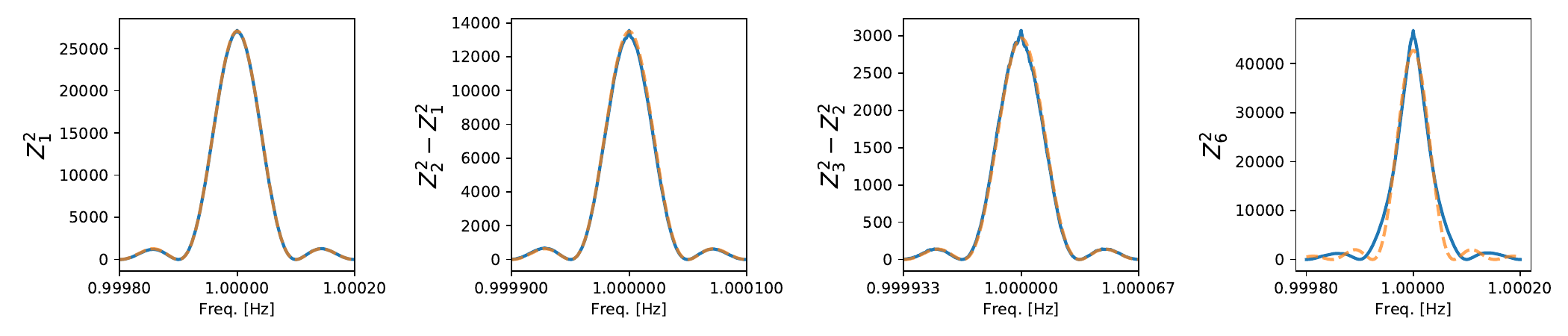}}\\
    {\includegraphics[width=0.95\textwidth]{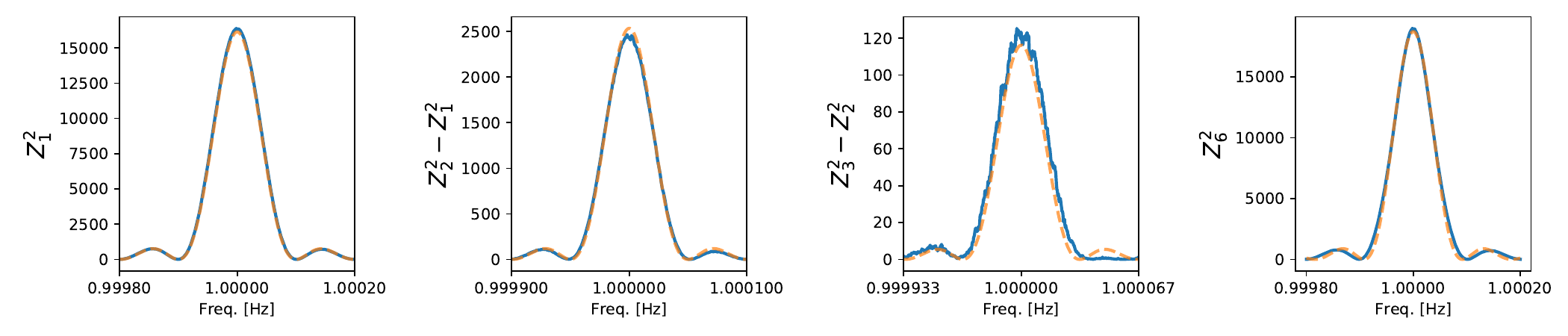}}
    \caption{\textit{Top to bottom:} The first three column shows a comparison of measured quantity of $Z_1^2, (Z_2^2-Z_1^2)$ and $(Z_3^2-Z_2^2)$ [Blue solid line] against the overlay of theoretically expected sinc-squared functions as per  [orange dashed line], for the simulated ToA of rectangular and Gaussian light curve, respectively.  In the last column, the best curve fit of a sinc-squared function is plotted over the measure value of $Z_6^2$ for both the light curves.  }
    \label{fig:plot_array}
\end{figure*}

\begin{table}
	\centering
	\caption{Grid of source parameters (column~1) which were randomly uniformly sampled in log-scale, within the range of column 2 and columns 3,  with number of samples mentioned in column~4. Since $b$ can not be arbitrarily sampled, it was constrained to within the range of $0.1a$ to $0.5a$ (as per the sampled $a$). A large dynamic parameter space for used to avoid any selection biases.\label{tab:grid_composition2}}
	\begin{tabular}{lccr} 
		\hline
		Parameter & Lower Bound & Upper Bound  & Number\\
		\hline
		$f_0~[Hz]$ & 0.1 & 100 &   20\\
		$a~[s^{-1}]$ & 35 & 350 & 20\\
		$b~[s^{-1}]$ & $0.1a$ & $0.5a$ &  20\\
		$T~[s]$ & 30 & 120000 & 20\\
		\hline
	\end{tabular}
\end{table}

\begin{figure*}
\centering
  \includegraphics[width=0.9\textwidth]{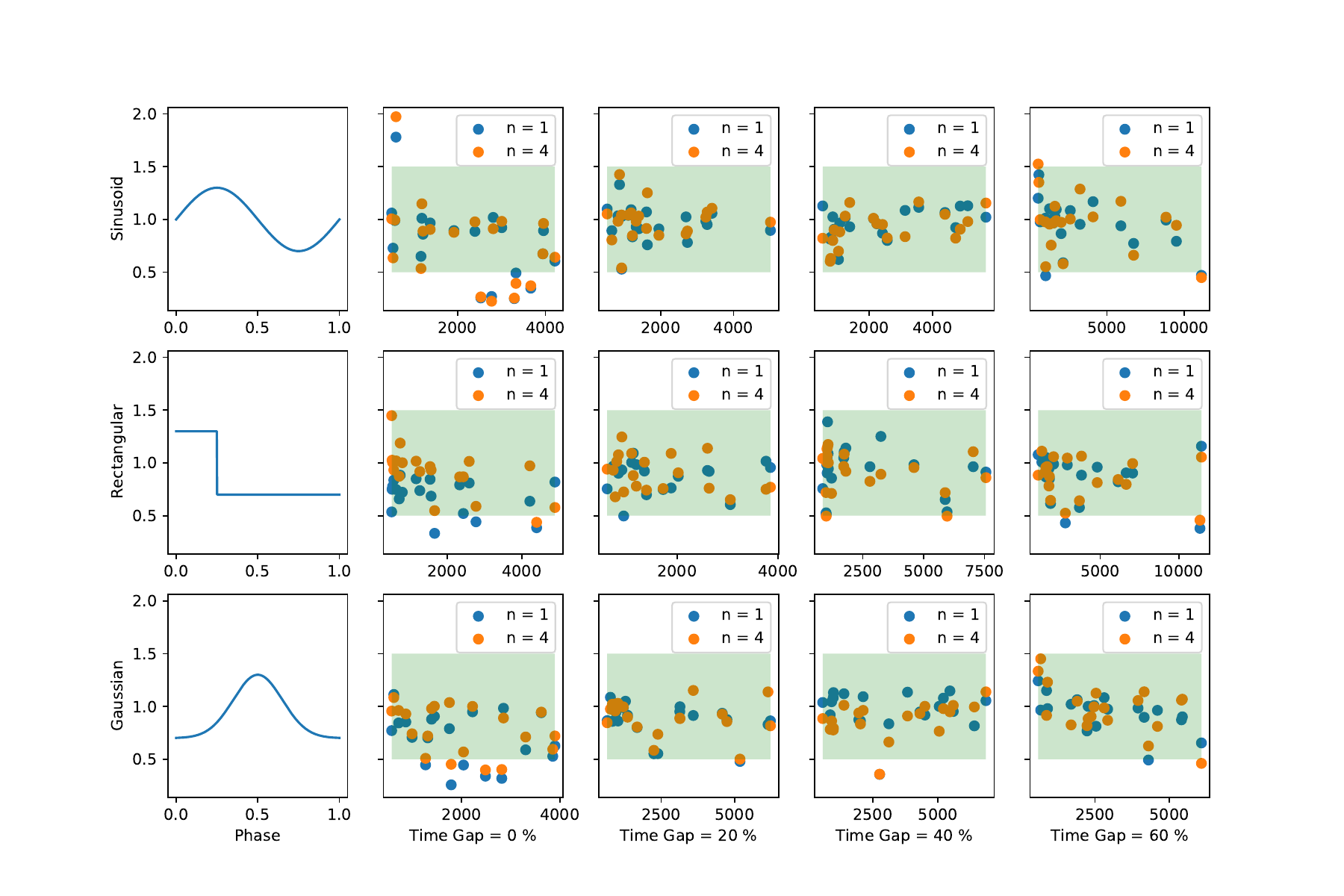}
  \caption{\label{real_AWlawrobust} 
  Error ratios plotted for real data from three pulsars with light curve shapes, (\textit{top to bottom}).  The ER were computed for 3 different fractions of time gap, (\textit{left to right}) 0, 20, 40 and 60 per cent respectively. Under all these scenarios the Error.}
 
\end{figure*}

The $AW$ power-law in Eq.~(\ref{AWpowerlaw}) has only been tested for sinusoidal simulations, where contribution from higher harmonics is insignificant. Therefore, it is necessary to investigate the behavior of the power-law for different light curves.

We choose rectangular and a Gaussian shaped pulse, to study the behaviour of the power-law in the presence of the higher harmonics, and define a pulse of length 1 second as,
\begin{align}
y_{\rm rect}(t; \rm{width}^{\prime}) &= \begin{cases}
    a+b,& \text{if } 0\leq t < \rm{width^{\prime}}~\\
    a,              & \rm{width}^{\prime} \leq t < 1
\end{cases},\label{RECT}\\
y_{\rm gauss}(t; \rm{width}^{\prime}) &= a + b\exp\left[{-\left(\frac{t-1/2  }{0.25~\rm{width}^{\prime}}\right)^2}\right].\label{GAUSS}
\end{align}
We have already defined $a$ and $b$ in Sec.~\ref{lightcurve} and the `width$^{\prime}$' dictates the fraction of time the pulse is above its minima. It is noteworthy that the defined $y_{\rm gauss}$ is a truncated Gaussian function, which presents a non-differentiable point at the intersection of two pulses. However, these will have insignificant impact on the simulations. 
Based on these definitions, we can extend it as a periodic light curve using the condition $y(t +1) = y(t)$, and can then generate a light curve with frequency $f_0$ and initial phase $\Phi$ using the function $y(f_0t + \Phi)$. For our simulation we chose $y_{\rm rect}(t; 0.25)$ and  $y_{\rm gauss}(t; 0.5)$. The simulated ToA and their Fourier components in expansion as per Eq.~(\ref{fourier}) are shown in Fig.~\ref{fig:shape_toa}. Since Fourier components diminish after fifth harmonics, we limit the simulations till $Z_6^2$. Fig.~\ref{fig:plot_array} validates Eq.~(\ref{harm}) for $k \in \{1,2,3\}$ and also justifies the assumption that $Z_6^2$ (which is a sum of sinc-squared terms) can be well approximated as one sinc-squared term in small region near $f_0$ for both $y_{\rm rect}$ and $y_{\rm gauss}$ light curve.

In order to compare the MAD$(f_r)$ with the $AW$ power-law, the same methodology was employed. The term `ER' in Eq.~(\ref{error_ratio}) has been redefined by substituting SMAD with power-law from Eq.~(\ref{AWpowerlaw}) as,
 \begin{align}
     \rm{ER} &= \frac{\text{Numerically computed MAD($f_r$)}}{W_r\sqrt{(1.5A_r^{-1})} }.\label{ER_new}
 \end{align}
This approach aims to further investigate the efficacy of the proposed measure by comparing it with a previously established power-law. 

We proceed with simulations by  sampling the source parameters in log-scale as described in Table~\ref{tab:grid_composition2}. Lastly we introduce gaps in the simulation to mimic gaps found in the real X-ray data, which we explore in the next section.

\subsection{Robustness to Time Gaps}
\begin{figure}
\centering
  \includegraphics[width=0.45\textwidth]{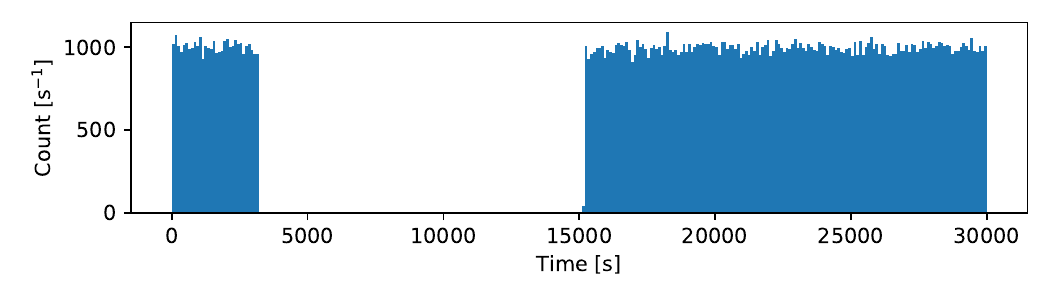}
  \includegraphics[width=0.45\textwidth]{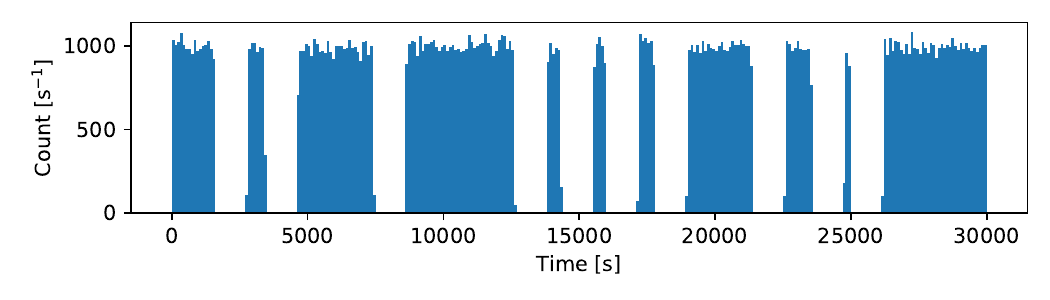}
  \caption{Binned ToA, of which 40 per cent of the data is removed by two methods, \textit{top:} a contiguous gap, randomly introduced within the time range of [0.05$T$ -- 0.95$T$] of the observation and, b) \textit{bottom:} where ten non overlapping gaps of 4 per cent each and randomly introduced in the data.}
  \label{gap_multi}
\end{figure}

\begin{table}
	\centering
	\caption{Comparison of MAD and ER for 
three differently shaped light curve with source parameters $(a, b, f_0, T) = (10,5,1,1000)$, as described in Eqs.~(\ref{sinecurve}), (\ref{RECT}) and~(\ref{GAUSS}). The data have been subjected to different gaps of $\{20,40, 60\}$ per cent for all shapes and no gaps for rectangular and Gaussian light curves. This is a qualitative representation of variation in ER in different regimes. \label{tab:robust_shapes_gaps}}
	\begin{tabular}{lp{1cm}ccccr} 
		\hline
		Shape & Time Gap ($\%$) & $n$ & \multicolumn{2}{c}{Contiguous gap} & \multicolumn{2}{c}{Multiple gaps} \\
  &&&MAD&ER &MAD&ER \\
		\hline
		\multirow{6}{*}{Sine} 
		 & \multirow{2}{*}{20} & 1 & $8.4 \times 10^{-6}$ &   0.7 & $1.8 \times 10^{-5}$ &   1.4 \\
		 &  & 6 & $9.4 \times 10^{-6}$ &   0.8 & $1.7 \times 10^{-5}$ &   1.4 \\[0.5ex]
		 & \multirow{2}{*}{40} & 1 & $1.3 \times 10^{-5}$ &   0.9 & $1.6 \times 10^{-5}$ &   1.2 \\
		 &  & 6 & $1.3 \times 10^{-5}$ &   0.9 & $1.9 \times 10^{-5}$ &   1.4 \\[0.5ex]
		 & \multirow{2}{*}{60} & 1 & $1.0 \times 10^{-5}$ &   0.5 & $1.7 \times 10^{-5}$ &   1.0 \\
		 &  & 6 & $9.0 \times 10^{-6}$ &   0.5 & $1.2 \times 10^{-5}$ &   0.8 \\[0.5ex]
		\hline
		\multirow{8}{*}{Rect} 
          & \multirow{2}{*}{0} & 1 & $6.6 \times 10^{-6}$ &   0.8 & -- & -- \\
		 &  & 6 & $5.8 \times 10^{-6}$ &   0.8 & -- & -- \\[0.5ex]
		 & \multirow{2}{*}{20} & 1 & $8.1 \times 10^{-6}$ &   0.8 & $5.1 \times 10^{-6}$ &   0.5 \\
		 &  & 6 & $7.1 \times 10^{-6}$ &   0.8 & $5.4 \times 10^{-6}$ &   0.6 \\[0.5ex]
		 & \multirow{2}{*}{40} & 1 & $9.3 \times 10^{-6}$ &   0.7 & $9.9 \times 10^{-6}$ &   0.9 \\
		 &  & 6 & $6.1 \times 10^{-6}$ &   0.5 & $9.4 \times 10^{-6}$ &   1.0  \\[0.5ex]
		 & \multirow{2}{*}{60} & 1 & $1.3 \times 10^{-5}$ &   0.8 & $7.6 \times 10^{-6}$ &   0.6 \\
		 &  & 6 & $6.3 \times 10^{-6}$ &   0.4  & $6.3 \times 10^{-6}$ &   0.6 \\[0.5ex]
		\hline
		\multirow{8}{*}{Gauss}
        & \multirow{2}{*}{0} & 1 & $1.5 \times 10^{-5}$ &   0.6 & -- & -- \\ 
		 &  & 6 & $1.7 \times 10^{-5}$ &   0.8 & -- & -- \\[0.5ex]
		 & \multirow{2}{*}{20} & 1 & $2.3 \times 10^{-5}$ &   0.8 & $3.1 \times 10^{-5}$ &   1.1 \\
		 &  & 6 & $2.0 \times 10^{-5}$ &   0.8 & $2.3 \times 10^{-5}$ &   1.0 \\[0.5ex]
		 & \multirow{2}{*}{40} & 1 & $1.9 \times 10^{-5}$ &   0.6 & $2.3 \times 10^{-5}$ &   0.7 \\
		 &  & 6 & $2.0 \times 10^{-5}$ &   0.6 & $2.9 \times 10^{-5}$ &   1.0 \\[0.5ex]
		 & \multirow{2}{*}{60} & 1 & $3.0 \times 10^{-5}$ &   0.7 & $8.5 \times 10^{-6}$ &   0.2 \\
		 &  & 6 & $2.9 \times 10^{-5}$ &   0.7 & $1.3 \times 10^{-5}$ &   0.4 \\[0.5ex]
		\hline
	\end{tabular}
\end{table}

The real astronomical data consists of data gaps because of various reasons. 
These data with gaps may not follow some of the assumptions we have used in Sec.~\ref{sec_analytical} and can introduce unwanted artefacts in our methodology. Hence, we investigate the robustness of $AW$ power-law derived in Sec.~\ref{varomega} on real X-ray data, following the same procedure described in Sec~\ref{multiple_simulations}. We have introduced the two kinds of gaps (Fig.~\ref{gap_multi}); i) a contiguous long gap, ii) 10 separate gaps of the same size in various parts of the data. \par 
For ToA of observation time $T$, the gaps are introduced only within the time range [0.05$T$ -- 0.95$T$] to ensure that the time difference between the first and last photon of ToA is always  $T$. The gap(s) were chosen to be $\{20,40,60\}$ per cent of the observation time.

Table~\ref{tab:robust_shapes_gaps} shows the comparisons in the errors and ER measured for different shapes and gaps for 1 set of parameters $(a,b,f_0, T) = (10,5,1,1\,000)$. Furthermore, Fig.~\ref{real_AWlawrobust} shows ER for the parameters mentioned in Table~\ref{tab:grid_composition2}. Since we have not imposed any limitation on `SS' or on the number of cycles of the signal, the ER is conservatively within the factor of 2.
However, the ER is still robust within acceptable tolerances for high `SS' signals. Although the behavior of $Z_n^2$ in the time gap regime can be delved into more analytically, as detailed in Sec.~\ref{sec:time--gaps}, our initial findings suggest that the approach presented here is simpler and more reliable.

\begin{table}
\centering
\caption{Comparison of MAD and ER using \emph{Crab} data for various observation times and most contributing four harmonics $n\in\{1,2,3,4\}$.   
\label{tab:robust_real_data}}
\begin{tabular}{p{1.2cm}cccccccr} 
\hline
Time [s] ($T$) & $n$ & MAD & ER& Freq [Hz]\\
\hline
\multirow{ 4}{*}{50} & 1 & $3.9\times 10^{-4}$ & 0.9 & 26.655403 \\
  & 2 & $1.5\times 10^{-4}$ & 0.9 & 26.655389\\
  & 3 & $1.0\times 10^{-4}$ & 1.0 & 26.655379\\
  & 4 & $1.1\times 10^{-4}$ & 1.1 & 26.655371\\
\hline
  \multirow{ 4}{*}{100} & 1 & $1.6\times 10^{-4}$ & 1.1 & 26.655410\\
  & 2 & $7.5\times 10^{-4}$ & 1.2 & 26.655405\\
  & 3 & $5.1\times 10^{-5}$ & 1.4 & 26.655413\\
  & 4 & $6.3\times 10^{-5}$ & 1.8 & 26.655412\\
\hline
\end{tabular}
\end{table}

\subsection{Test with \emph{Crab} pulsar\label{astrosat_data}}

\begin{figure*}
\centering
  \includegraphics[width=1.0\textwidth]{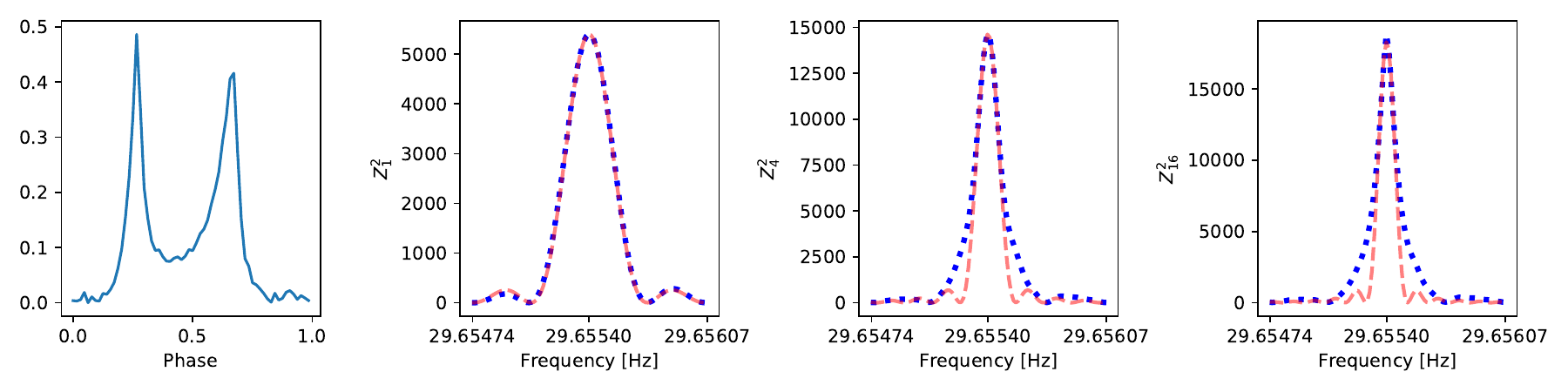}
  \caption{The first plot shows the phase folded light curve of the \emph{Crab} using \emph{LAXPC} data. From second to the fourth plot, $Z_n^2$ statistics is plotted for the observation time 3000~s for three different harmonics $n \in {1,4,16}$, respectively.   This shows that the $Z_n^2$ can be approximated by a sinc-squared, even for higher harmonics for generic shape pulse profile as well.}
  \label{goodsinc}
\end{figure*}

Unlike simulated data, the pulsar data contains ill behaved non-stationary noise, irregular time gaps and most importantly the pulsar signal need not be a constant frequency signal. So far in all our simulations we have assumed $f_0$ is constant.

In this section, we demonstrate the validity of our $AW$ power-law on observational data of \emph{Crab} pulsar.  We have used the data from \emph{AstroSat}'s \citep{2014_sing_AstroSat} Large Area X-ray Proportional Counters [\emph{LAXPC} (3.0-80.0 keV)], because of its excellent time resolutions. 
We have used the LAXPC data of \emph{Crab} with Obs Id 1876, to test our methodology.
For \emph{LAXPC}, we have downloaded the level 1 data from the archive\footnote{\url{https://astrobrowse.issdc.gov.in/astro_archive/archive/Home.jsp}}. We have used the \texttt{laxpcsoft} package available at the \emph{AstroSat} Science Support Cell (ASSC)\footnote{\url{http://astrosat-ssc.iucaa.in}} for our analysis. We have created the event files using the standard task \texttt{laxpc\_make\_event}. After that, we have used \texttt{laxpc\_make\_stdgti} to generate a good time interval (\textit{gti}) file, removing intervals of passage through the South Atlantic Anomaly (SAA) and Earth occultation. To improve the signal to noise ratio (SNR), we have selected the events only from \emph{LAXPC} 10, in the energy range 3-25 keV (layer 1 of \emph{LAXPC}) which is also consisted with the \texttt{gti}. {The application of these filtrations removes $\sim60$ per cent of the data. This also justifies the choice of time gaps in the previous section.} Furthermore, as it is difficult to ensure all realisations have equivalent gaps and background noise, we limit our investigation to smaller $T\sim \mathcal{O}(100~\rm{s})$ within a gap-less and small time interval $\sim 3000~\rm{s}$.

The frequency of the X-ray signals from \emph{Crab} pulsar decreases with time at the rate $\dot{f}\sim 10^{-10}$~[Hz~s$^{-1}$] \citep{1988MNRAS.233..667L}. Since our investigation involves smaller observation time \textit{i.e.,} $T^2 \ll 1/\dot{f}$, the effects due to change in frequency can be ignored.
We find that for real \emph{Crab} data, the $Z_n^2$ distribution can be approximated to a sinc-squared function in a small region around the source frequency. In Fig.~\ref{goodsinc} we have shown the distribution of $Z_n^2$ statistics for the \emph{Crab} data for three different harmonics $n \in \{1,4,16\}$.   

In order to calculate MAD($f_r$), we use an observational data of 3\,000 s and divide it into 300 overlapping observational sections of $T\in\{50,100\}$ second. { We then compute the MAD and ER in the real data regime using our methodology. The  Table~\ref{tab:robust_real_data}, shows ER to be consistent within the error margin, and $AW$ power-law is reasonably reliable in real pulsar regime.} The observed deviation in the higher harmonics over a 100~s observation is likely attributable to the comparatively lower SNR of the higher harmonics and the consequential non-negligible influence of the spin-down terms. In the next section we compare how our methodology is compared to other formulations commonly used in the literature.

\section{Comparison With other error estimation Methods}\label{sec_comparewithothers}

\begin{figure}
\centering
  \includegraphics[width=0.5\textwidth]{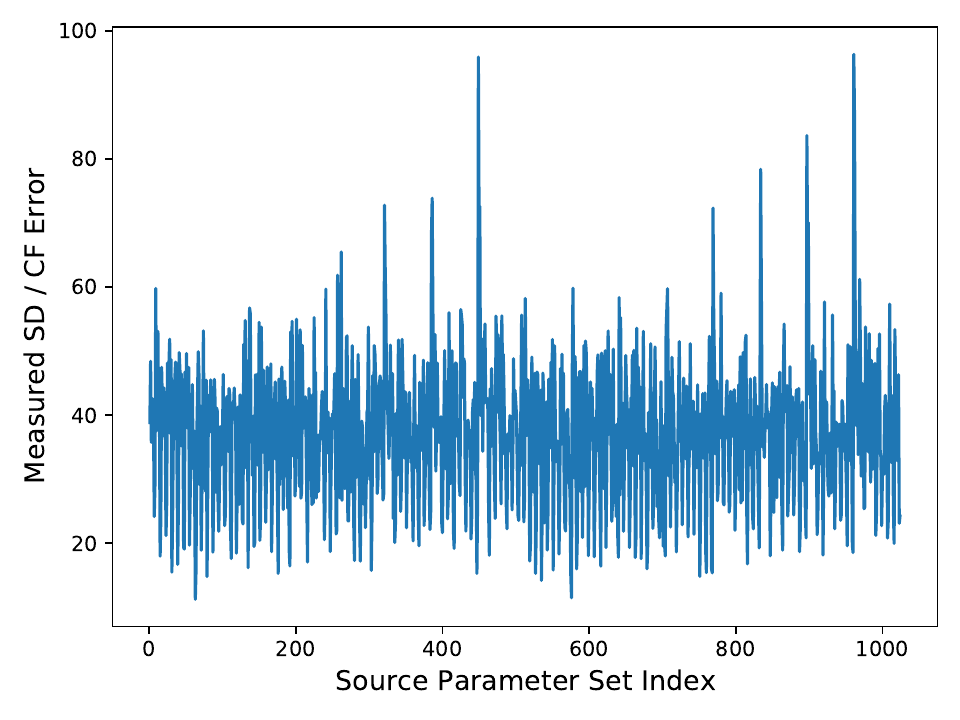}
  \caption{Ratio of measured standard deviation (SD) from the distribution of frequencies  to the standard deviation in the curve fitting (CF) frequency output by the fitting routine. Here the parameters of sinusoidal pulsars are same as a described in Table.~\ref{tab:grid_composition}}
  \label{CEMADratios}
\end{figure}

Methods widely used in literature (\textit{e.g.,}  \citep{Bloomfield1976,Kovacs1980,Larsson1996}) rely only on the variance of the curve fitting parameters for estimating the uncertainty in frequency and end up under-estimating it, as shown in Eq.~(\ref{varfr}). The Fig.~\ref{CEMADratios} demonstrates that the standard deviation in fitting parameters is often orders of magnitude smaller than the standard deviation of the recovered frequencies. Moreover, the curve fitting error exhibits dependency on the number of data points used in the routine. This makes the error dependent on users' choice rather than being purely dependent on the data and the source parameters. In our study we have found MAD to be robust towards the number of data points used to compute the $Z_n^2$ statistics. 

Fitting-based methodologies assume that the data is a time series governed by a sinusoidal (or sum of sinusoids) light curve with a time-independent Gaussian noise having a standard deviation of $\sigma_{\rm{total}}$. However, high-energy pulsar data has discrete times of arrival (ToA), and only after binning the ToA (and hence losing information) do we get a time series with time-dependent Poisson noise. As discussed in \citep{Bloomfield1976}, the standard deviation in fitting the angular frequency ($\omega = 2\pi f_0$) is given by,
   
    \begin{align}
        \sigma_{\omega}^{*} = \frac{\sqrt{24}\:\sigma_{\rm{tot}}}{N_d^{1.5}\: b},\label{bloomfield_error}
    \end{align}
where $N_d$ is the number of data points and $b$ is the amplitude of the sinusoidal source. Rewriting Eq.~(\ref{bloomfield_error}) in terms of the variables used in this study, we can approximate $\sigma_{\rm{tot}} \sim \sqrt{a}$ and $N_d = T/\Delta~t$. Thus, we get,
\begin{align}
    \sigma_{\omega}^{*} = \sigma^{*}({2\pi f_r}) \sim \sqrt{24}~\frac{\sqrt{a}}{b T^{1.5}} \Delta t^{1.5}.\label{bloomfield_to_AI}
\end{align}
    This formulation is similar to Eq.~(\ref{AWpowerlaw}) except for an additional factor proportional to $\Delta t^{1.5}$. As in typical $\Delta t \lesssim 1/(10f_0)$, we underestimate the error by a few orders of magnitude. Similar formulations are discussed in \citep{Kovacs1980} and  for the PIPS pipeline in \citep{2021arXiv210714223M,2001AIPC..568..557G}. By extending the same line of reasoning, it can be shown that they under-estimate the error by a factor of
    $\gtrsim\sqrt{\Delta t}$. 

It is to be noted that the error estimates discussed in this section so far assign appropriate confidence intervals in fitting parameters for the relevant data type (and within their assumptions). However, these confidence intervals can not be extended to the source frequency of the signal without taking the contribution of the Poisson effects in Poisson limited periodic data.   

The methods employed by \citet{2005ApJ...618..866N, 2017NewA...56...94B} to measure uncertainties in pulse periods at the 1$\sigma$ confidence level involve period range where the $\chi^2$ decreases from its peak by one standard deviation of $\chi^2$ values over an extended period range. Such methods might lead to a larger estimation of the error. To understand this, we relate the Width $W$ of the $Z_1^2$ sinc-squared statistic to the standard deviation $\sigma$, under an assumption of an approximate Gaussian like behaviour around the peak. The relationship is depicted as $W = \sigma/1.22$, and can be further elaborated by Taylor expansion around the peak as,

\begin{align}
A - \frac{A}{3W^2}(f-f_0)^2 &= A - \frac{A}{2 \sigma^2}(f - f_0)^2\\
\Rightarrow \sigma &= \sqrt{\frac{3}{2}}W
\end{align}

Substituting the standard deviation with Full Width at Half Maxima (FWHM) doesn't resolve the estimation concern, given $W = \text{FWHM}/2.78$. Hence, it can be concluded that error estimation by these methods can be expressed as,

\begin{align}
    \sigma^{*}({f_r}) \sim W.
\end{align}

Using Eq.~(\ref{AWpowerlaw}), we see that this over-estimates the error by a factor of $\sqrt{A}$. Since typically $A \gg 1$, we might wrongly compute an error which is orders of higher. This over-estimation can not be negated by measuring the width at half the maxima 
(or at any fraction of the maxima), as this quantity is independent of the amplitude $A$ in Eq.~(\ref{AWpowerlaw}).  

The Monte Carlo method, as explained by \citet{2013AstL...39..375B}, shares similarities with the error methodology proposed in our study. In the Monte Carlo approach, a simulated light curve is generated for a large number of iterations ($n_{\rm{sim}}\simeq 1000$) and the standard deviation is calculated by fitting a Gaussian function to the epoch folding statistics. Our methodology improves upon this by eliminating the need for simulations, resulting in a more efficient approach for computing errors. In the next section we discuss the impact of improved error estimations on results based on both over and under-estimated errors.   

For sinusoidal time series influenced by white noise, \citep{ransom} demonstrated that the frequency error, $\sigma_f$, is given by:
\begin{equation}
    \sigma_f = \frac{3}{\pi\alpha\sqrt{6P(f_0)}}
\end{equation}
Where $P(f)$ is analogous to $Z_n^2(f)$ and the term $\alpha$ is defined as:
\begin{equation}
    \alpha = \frac{1}{\pi} \sqrt{-\frac{3}{2P(f_0)} \frac{\partial^2 P(f_0)}{\partial f^2}}
\end{equation}
This can be further simplified to:
\begin{equation}
    \sigma_f = \cfrac{1}{\sqrt{-\cfrac{\partial^2 P(f_0)}{\partial f^2}}}
\end{equation}
This is congruent with the findings of \citet{chang}, which are grounded in the arguments presented by \citet{bretthorst2013bayesian}, leading to the relation:
\begin{equation}
    \delta f = \frac{\pi\sqrt{3}}{T_{\text{obs}}}(Z^2_{1,\text{max}})^{-\frac{1}{2}}.
\end{equation}
However, these works focus on continuous time-domain signals. Our research establishes the applicability of these equations to Poisson-limited data over extended observation periods, and offers a rigorous method to estimate errors reliably.
 
\section{Application and Impact}\label{sec_impact}
Analysis involving the frequency of a signal often has to take into account the error in the measurement of the frequency. This error propagates to the parameters that are correlated with the frequencies, so errors can significantly affect our confidence intervals in models. Some examples are discussed below,
\begin{enumerate}
    \item Methods that underestimate the errors in frequency can lead to false modeling of fluctuations in the source frequency. Subsequent frequency measurements could be well outside the confidence interval of previously estimated errors.
    \item Pulsars frequencies can vary with time, and measurements of frequency derivatives are important in modeling pulsars and their interactions with the environment. Therefore, improper errors in frequency can propagate to spin-down terms and can significantly affect the analysis. 
    \item Directed pulsars with known ephemeris are prime science targets for gravitational wave (GW) searches. Based on commonly accepted general relativity theory the frequency of the GW from pulsars is twice the spin frequency. Hence the error region around the spin frequency determines the search region of the GW search pipelines . Computationally expensive searches as described in  \citet{2000PhRvD..61h2001B,2010PhRvD..82d2002P, 2012PhRvD..85h4010P, 2019CQGra..36t5015S} will be largely affected if the error region changes significantly. It has already been shown in Sec.~\ref{sec_comparewithothers} that the error computed  using Eq.~(\ref{AWpowerlaw}) is orders of magnitude different from other estimates. Therefore,  this shift in computational burden can affect the upper limit and the confidence interval on several parameters associated in modeling of the gravitational wave sources.
\end{enumerate}

\begin{figure}
    \centering
\begin{tikzpicture}[node distance=2.5cm]
\node (start) [startstop, text width=4cm] {ToA from periodic Poisson limited data for observation time $T$.};
\node (in1) [process, below of=start, text width=5cm, node distance=2.0cm] {Compute the $Z_n^2$ statistics around the rough frequency estimate for a range $2/T$ with appropriate resolution,
};
\node (pro1) [process, text width=5cm, below of=in1] {Identify the frequency $f_{\rm max}$ where $Z_n^2$ is maximum and use statistics for frequencies within $f_{\rm max} \pm 2/(T\pi)$ .};
\node (pro2) [process, text width=5cm, below of=pro1] {Obtain the sinc-squared parameters $A_r$, $f_r$ and $W_r$ by curve fitting within the range of $2/(T\pi)$ .};
\node (pro3) [process, below of=pro2, text width=5cm] {The error associated with the $f_r$ of signal is given with $\sim 68$ per cent confidence be $W_r\sqrt{3}/\sqrt{A_r}$.};
\draw [arrow] (start) -- (in1);
\draw [arrow] (in1) -- (pro1);
\draw [arrow] (pro1) -- (pro2);
\draw [arrow] (pro2) -- (pro3);
\end{tikzpicture}

\caption{Flow chart of the recipe for determining the spin frequency of a pulsar and the associated confidence interval for a given times of arrivals from a pulsar.\label{flow2}}    
\end{figure}
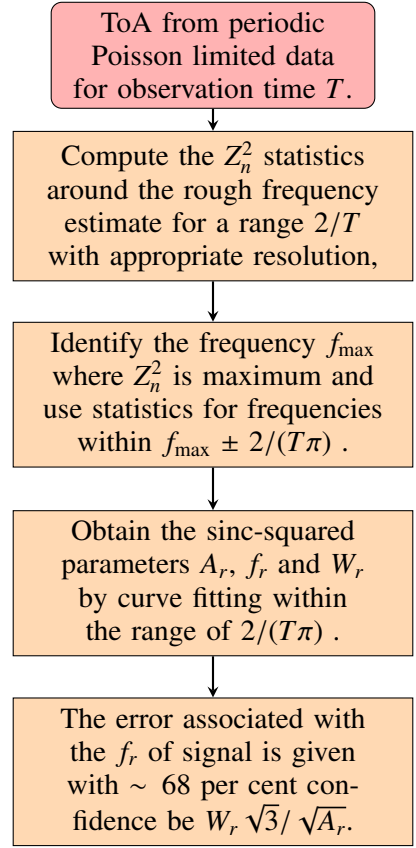

\section{Conclusion}\label{sec:conclusion}
High energy time-tagged photon data are often used for determining the frequency of a period signal, such as X-ray pulsars. Accurately estimating the uncertainties in the inferred frequency is important for any subsequent analyses using those values. Unfortunately, there is no widely accepted method, leading to wrong uncertainties being reported in some cases.

We present a novel methodology for determining the uncertainty in frequency of pulsar-like periodic signal in a Poisson-limited data, using a reliable and trivial approach for computing errors with appropriate confidence intervals. Our methodology provides a robust solution that does not require computationally expensive simulations, offering a light-curve independent approach that improves the accuracy of frequency estimates. We provide a flow chart summarizing our methodology in Fig.~\ref{flow2} for reasonably well-behaved signals. Potential enhancements and fine-tuning of the curve fitting process are further explored in Sec.~\ref{sec:finetune}

We present a detailed mathematical reasoning for the contribution of different components in determining the uncertainty.
We demonstrated the efficacy of our approach by extensively testing it on a variety of simulated data sets including data with gaps, and observational data from the \emph{Crab} pulsar. 
We compared our methodology with existing methods from the literature and found that our approach predicts errors with significantly improved confidence intervals. Our findings reveal that existing methods can often be an order of magnitude off, underscoring the importance of a reliable and computationally efficient method for periodic data analysis.  Our methodology shares the limitation of the $Z_n^2$ search and assumes the variation in the frequency to be statistically insignificant within the observation time. Future studies could investigate how these limitations may be overcome, and also how the error in frequency propagates to the spin-down terms. 

\section*{Acknowledgements}\label{sec_acknowledgements}
We are thankful to \emph{AstroSat} Science Support Cell and ISRO Science Data Archive for \emph{AstroSat} Mission for providing the necessary data and pipeline for studying the \emph{Crab} data.  We express our heartfelt appreciation to Santosh~Vadawale, A.~R.~Rao, Deepto Chakrabarty, and Dipankar Bhattacharya 
for providing us with invaluable insights on the topic. Their guidance has been crucial in shaping our understanding. 
This analysis extensively used Python and various libraries including \texttt{Stringray} \citep{stingray}, \texttt{Astropy} \citep{astropy:2013, astropy:2018, astropy:2022} and \texttt{Scipy} \citep{2020SciPy-NMeth}. We also used the \emph{AstroSat/LAXPC} data analysis pipeline.\\

\section*{Data Availability}\label{sec_data}
The pipeline developed for this study can be found in \url{https://github.com/akshats14/frequency_error_pulsar/}. For raw data of \emph{Crab} pulsar used in our investigation, please send a request to the authors.

\bibliography{ref}

\begin{thebibliography}{}
\expandafter\ifx\csname natexlab\endcsname\relax\def\natexlab#1{#1}\fi

\bibitem[{{Astropy Collaboration} {$et~al$.}(2013){Astropy Collaboration},
  {Robitaille}, {Tollerud}, {Greenfield}, {Droettboom}, {Bray}, {Aldcroft},
  {Davis}, {Ginsburg}, {Price-Whelan}, {Kerzendorf}, {Conley}, {Crighton},
  {Barbary}, {Muna}, {Ferguson}, {Grollier}, {Parikh}, {Nair}, {Unther},
  {Deil}, {Woillez}, {Conseil}, {Kramer}, {Turner}, {Singer}, {Fox}, {Weaver},
  {Zabalza}, {Edwards}, {Azalee Bostroem}, {Burke}, {Casey}, {Crawford},
  {Dencheva}, {Ely}, {Jenness}, {Labrie}, {Lim}, {Pierfederici}, {Pontzen},
  {Ptak}, {Refsdal}, {Servillat}, \& {Streicher}}]{astropy:2013}
{Astropy Collaboration}, {Robitaille}, T.~P., {Tollerud}, E.~J., {$et~al$.}
  2013, \aap, 558, A33

\bibitem[{{Astropy Collaboration} {$et~al$.}(2018){Astropy Collaboration},
  {Price-Whelan}, {Sip{\H{o}}cz}, {G{\"u}nther}, {Lim}, {Crawford}, {Conseil},
  {Shupe}, {Craig}, {Dencheva}, {Ginsburg}, {Vand erPlas}, {Bradley},
  {P{\'e}rez-Su{\'a}rez}, {de Val-Borro}, {Aldcroft}, {Cruz}, {Robitaille},
  {Tollerud}, {Ardelean}, {Babej}, {Bach}, {Bachetti}, {Bakanov}, {Bamford},
  {Barentsen}, {Barmby}, {Baumbach}, {Berry}, {Biscani}, {Boquien}, {Bostroem},
  {Bouma}, {Brammer}, {Bray}, {Breytenbach}, {Buddelmeijer}, {Burke},
  {Calderone}, {Cano Rodr{\'\i}guez}, {Cara}, {Cardoso}, {Cheedella}, {Copin},
  {Corrales}, {Crichton}, {D'Avella}, {Deil}, {Depagne}, {Dietrich}, {Donath},
  {Droettboom}, {Earl}, {Erben}, {Fabbro}, {Ferreira}, {Finethy}, {Fox},
  {Garrison}, {Gibbons}, {Goldstein}, {Gommers}, {Greco}, {Greenfield},
  {Groener}, {Grollier}, {Hagen}, {Hirst}, {Homeier}, {Horton}, {Hosseinzadeh},
  {Hu}, {Hunkeler}, {Ivezi{\'c}}, {Jain}, {Jenness}, {Kanarek}, {Kendrew},
  {Kern}, {Kerzendorf}, {Khvalko}, {King}, {Kirkby}, {Kulkarni}, {Kumar},
  {Lee}, {Lenz}, {Littlefair}, {Ma}, {Macleod}, {Mastropietro}, {McCully},
  {Montagnac}, {Morris}, {Mueller}, {Mumford}, {Muna}, {Murphy}, {Nelson},
  {Nguyen}, {Ninan}, {N{\"o}the}, {Ogaz}, {Oh}, {Parejko}, {Parley}, {Pascual},
  {Patil}, {Patil}, {Plunkett}, {Prochaska}, {Rastogi}, {Reddy Janga},
  {Sabater}, {Sakurikar}, {Seifert}, {Sherbert}, {Sherwood-Taylor}, {Shih},
  {Sick}, {Silbiger}, {Singanamalla}, {Singer}, {Sladen}, {Sooley},
  {Sornarajah}, {Streicher}, {Teuben}, {Thomas}, {Tremblay}, {Turner},
  {Terr{\'o}n}, {van Kerkwijk}, {de la Vega}, {Watkins}, {Weaver}, {Whitmore},
  {Woillez}, {Zabalza}, \& {Astropy Contributors}}]{astropy:2018}
{Astropy Collaboration}, {Price-Whelan}, A.~M., {Sip{\H{o}}cz}, B.~M.,
  {$et~al$.} 2018, \aj, 156, 123

\bibitem[{{Astropy Collaboration} {$et~al$.}(2022){Astropy Collaboration},
  {Price-Whelan}, {Lim}, {Earl}, {Starkman}, {Bradley}, {Shupe}, {Patil},
  {Corrales}, {Brasseur}, {N{"o}the}, {Donath}, {Tollerud}, {Morris},
  {Ginsburg}, {Vaher}, {Weaver}, {Tocknell}, {Jamieson}, {van Kerkwijk},
  {Robitaille}, {Merry}, {Bachetti}, {G{"u}nther}, {Aldcroft},
  {Alvarado-Montes}, {Archibald}, {B{'o}di}, {Bapat}, {Barentsen}, {Baz{'a}n},
  {Biswas}, {Boquien}, {Burke}, {Cara}, {Cara}, {Conroy}, {Conseil}, {Craig},
  {Cross}, {Cruz}, {D'Eugenio}, {Dencheva}, {Devillepoix}, {Dietrich},
  {Eigenbrot}, {Erben}, {Ferreira}, {Foreman-Mackey}, {Fox}, {Freij}, {Garg},
  {Geda}, {Glattly}, {Gondhalekar}, {Gordon}, {Grant}, {Greenfield}, {Groener},
  {Guest}, {Gurovich}, {Handberg}, {Hart}, {Hatfield-Dodds}, {Homeier},
  {Hosseinzadeh}, {Jenness}, {Jones}, {Joseph}, {Kalmbach}, {Karamehmetoglu},
  {Ka{l}uszy{'n}ski}, {Kelley}, {Kern}, {Kerzendorf}, {Koch}, {Kulumani},
  {Lee}, {Ly}, {Ma}, {MacBride}, {Maljaars}, {Muna}, {Murphy}, {Norman},
  {O'Steen}, {Oman}, {Pacifici}, {Pascual}, {Pascual-Granado}, {Patil},
  {Perren}, {Pickering}, {Rastogi}, {Roulston}, {Ryan}, {Rykoff}, {Sabater},
  {Sakurikar}, {Salgado}, {Sanghi}, {Saunders}, {Savchenko}, {Schwardt},
  {Seifert-Eckert}, {Shih}, {Jain}, {Shukla}, {Sick}, {Simpson},
  {Singanamalla}, {Singer}, {Singhal}, {Sinha}, {Sip{H{o}}cz}, {Spitler},
  {Stansby}, {Streicher}, {{{S}}umak}, {Swinbank}, {Taranu}, {Tewary},
  {Tremblay}, {Val-Borro}, {Van Kooten}, {Vasovi{'c}}, {Verma}, {de Miranda
  Cardoso}, {Williams}, {Wilson}, {Winkel}, {Wood-Vasey}, {Xue}, {Yoachim},
  {Zhang}, {Zonca}, \& {Astropy Project Contributors}}]{astropy:2022}
{Astropy Collaboration}, {Price-Whelan}, A.~M., {Lim}, P.~L., {$et~al$.} 2022,
  apj, 935, 167

\bibitem[{{Bachetti} \& {Huppenkothen}(2022)}]{2022arXiv220907954B}
{Bachetti}, M., \& {Huppenkothen}, D. 2022, arXiv e-prints, arXiv:2209.07954

\bibitem[{{Bachetti} {$et~al$.}(2021){Bachetti}, {Pilia}, {Huppenkothen},
  {Ransom}, {Curatti}, \& {Ridolfi}}]{Bachetti}
{Bachetti}, M., {Pilia}, M., {Huppenkothen}, D., {$et~al$.} 2021, \apj, 909, 33

\bibitem[{{Beri} \& {Paul}(2017)}]{2017NewA...56...94B}
{Beri}, A., \& {Paul}, B. 2017, \na, 56, 94

\bibitem[{{Bloomfield}(1976)}]{Bloomfield1976}
{Bloomfield}, P. 1976, {Fourier analysis of time series: an introduction}

\bibitem[{{Boldin} {$et~al$.}(2013){Boldin}, {Tsygankov}, \&
  {Lutovinov}}]{2013AstL...39..375B}
{Boldin}, P.~A., {Tsygankov}, S.~S., \& {Lutovinov}, A.~A. 2013, Astronomy
  Letters, 39, 375

\bibitem[{{Brady} \& {Creighton}(2000)}]{2000PhRvD..61h2001B}
{Brady}, P.~R., \& {Creighton}, T. 2000, \prd, 61, 082001

\bibitem[{Bretthorst(2013)}]{bretthorst2013bayesian}
Bretthorst, G. 2013, Bayesian Spectrum Analysis and Parameter Estimation,
  Lecture Notes in Statistics (Springer New York)

\bibitem[{Bretthorst(1988)}]{Bretthorst1988}
Bretthorst, G.~L. 1988, Excerpts from Bayesian Spectrum Analysis and Parameter
  Estimation, ed. G.~J. Erickson \& C.~R. Smith (Dordrecht: Springer
  Netherlands), 75--145

\bibitem[{{Buccheri} {$et~al$.}(1983){Buccheri}, {Bennett}, {Bignami},
  {Bloemen}, {Boriakoff}, {Caraveo}, {Hermsen}, {Kanbach}, {Manchester},
  {Masnou}, {Mayer-Hasselwander}, {{\"O}zel}, {Paul}, {Sacco}, {Scarsi}, \&
  {Strong}}]{Buccheri1983}
{Buccheri}, R., {Bennett}, K., {Bignami}, G.~F., {$et~al$.} 1983, \aap, 128,
  245

\bibitem[{{Chang} {$et~al$.}(2012){Chang}, {Pavlov}, {Kargaltsev}, \&
  {Shibanov}}]{chang}
{Chang}, C., {Pavlov}, G.~G., {Kargaltsev}, O., \& {Shibanov}, Y.~A. 2012,
  \apj, 744, 81

\bibitem[{{Davies}(1990)}]{1990MNRAS.244...93D}
{Davies}, S.~R. 1990, \mnras, 244, 93

\bibitem[{{de Jager} \& {B{\"u}sching}(2010)}]{2010A&A...517L...9D}
{de Jager}, O.~C., \& {B{\"u}sching}, I. 2010, \aap, 517, L9

\bibitem[{{de Jager} {$et~al$.}(1989){de Jager}, {Raubenheimer}, \&
  {Swanepoel}}]{deJager1989}
{de Jager}, O.~C., {Raubenheimer}, B.~C., \& {Swanepoel}, J.~W.~H. 1989, \aap,
  221, 180

\bibitem[{{Gregory}(2001)}]{2001AIPC..568..557G}
{Gregory}, P.~C. 2001, in American Institute of Physics Conference Series, Vol.
  568, Bayesian Inference and Maximum Entropy Methods in Science and
  Engineering, ed. A.~{Mohammad-Djafari}, 557--568

\bibitem[{{Hare} {$et~al$.}(2021){Hare}, {Volkov}, {Pavlov}, {Kargaltsev}, \&
  {Johnston}}]{hare}
{Hare}, J., {Volkov}, I., {Pavlov}, G.~G., {Kargaltsev}, O., \& {Johnston}, S.
  2021, \apj, 923, 249

\bibitem[{Hart(1985)}]{hart1985choice}
Hart, J.~D. 1985, Journal of Statistical Computation and Simulation, 21, 95

\bibitem[{{Huppenkothen} {$et~al$.}(2019){Huppenkothen}, {Bachetti}, {Stevens},
  {Migliari}, {Balm}, {Hammad}, {Khan}, {Mishra}, {Rashid}, {Sharma}, {Martinez
  Ribeiro}, \& {Valles Blanco}}]{stingray}
{Huppenkothen}, D., {Bachetti}, M., {Stevens}, A.~L., {$et~al$.} 2019, \apj,
  881, 39

\bibitem[{{Kovacs}(1980)}]{Kovacs1980}
{Kovacs}, G. 1980, \apss, 69, 485

\bibitem[{{Larsson}(1996{\natexlab{a}})}]{1996A&AS..117..197L}
{Larsson}, S. 1996{\natexlab{a}}, \aaps, 117, 197

\bibitem[{{Larsson}(1996{\natexlab{b}})}]{Larsson1996}
---. 1996{\natexlab{b}}, \aaps, 117, 197

\bibitem[{{Leahy}(1987)}]{Leahy1987}
{Leahy}, D.~A. 1987, \aap, 180, 275

\bibitem[{{Leahy} {$et~al$.}(1983){Leahy}, {Elsner}, \&
  {Weisskopf}}]{Leahy1983b}
{Leahy}, D.~A., {Elsner}, R.~F., \& {Weisskopf}, M.~C. 1983, \apj, 272, 256

\bibitem[{{Lyne} {$et~al$.}(1988){Lyne}, {Pritchard}, \&
  {Smith}}]{1988MNRAS.233..667L}
{Lyne}, A.~G., {Pritchard}, R.~S., \& {Smith}, F.~G. 1988, \mnras, 233, 667

\bibitem[{{Murakami} {$et~al$.}(2021){Murakami}, {Jennings}, {Hoffman},
  {Sunseri}, {Baer-Way}, {Stahl}, {Savel}, {Altunin}, {Girish}, \&
  {Filippenko}}]{2021arXiv210714223M}
{Murakami}, Y.~S., {Jennings}, C., {Hoffman}, A.~M., {$et~al$.} 2021, arXiv
  e-prints, arXiv:2107.14223

\bibitem[{{Naik} {$et~al$.}(2005){Naik}, {Paul}, \&
  {Callanan}}]{2005ApJ...618..866N}
{Naik}, S., {Paul}, B., \& {Callanan}, P.~J. 2005, \apj, 618, 866

\bibitem[{{Nolan} {$et~al$.}(2012){Nolan}, {Abdo}, {Ackermann}, {Ajello},
  {Allafort}, {Antolini}, {Atwood}, {Axelsson}, {Baldini}, {Ballet},
  {Barbiellini}, {Bastieri}, {Bechtol}, {Belfiore}, {Bellazzini}, {Berenji},
  {Bignami}, {Blandford}, {Bloom}, {Bonamente}, {Bonnell}, {Borgland},
  {Bottacini}, {Bouvier}, {Brandt}, {Bregeon}, {Brigida}, {Bruel}, {Buehler},
  {Burnett}, {Buson}, {Caliandro}, {Cameron}, {Campana}, {Ca{\~n}adas},
  {Cannon}, {Caraveo}, {Casandjian}, {Cavazzuti}, {Ceccanti}, {Cecchi},
  {{\c{C}}elik}, {Charles}, {Chekhtman}, {Cheung}, {Chiang}, {Chipaux},
  {Ciprini}, {Claus}, {Cohen-Tanugi}, {Cominsky}, {Conrad}, {Corbet}, {Cutini},
  {D'Ammando}, {Davis}, {de Angelis}, {DeCesar}, {DeKlotz}, {De Luca}, {den
  Hartog}, {de Palma}, {Dermer}, {Digel}, {Silva}, {Drell}, {Drlica-Wagner},
  {Dubois}, {Dumora}, {Enoto}, {Escande}, {Fabiani}, {Falletti}, {Favuzzi},
  {Fegan}, {Ferrara}, {Focke}, {Fortin}, {Frailis}, {Fukazawa}, {Funk},
  {Fusco}, {Gargano}, {Gasparrini}, {Gehrels}, {Germani}, {Giebels},
  {Giglietto}, {Giommi}, {Giordano}, {Giroletti}, {Glanzman}, {Godfrey},
  {Grenier}, {Grondin}, {Grove}, {Guillemot}, {Guiriec}, {Gustafsson},
  {Hadasch}, {Hanabata}, {Harding}, {Hayashida}, {Hays}, {Hill}, {Horan},
  {Hou}, {Hughes}, {Iafrate}, {Itoh}, {J{\'o}hannesson}, {Johnson}, {Johnson},
  {Johnson}, {Johnson}, {Kamae}, {Katagiri}, {Kataoka}, {Katsuta}, {Kawai},
  {Kerr}, {Kn{\"o}dlseder}, {Kocevski}, {Kuss}, {Lande}, {Landriu},
  {Latronico}, {Lemoine-Goumard}, {Lionetto}, {Llena Garde}, {Longo},
  {Loparco}, {Lott}, {Lovellette}, {Lubrano}, {Madejski}, {Marelli}, {Massaro},
  {Mazziotta}, {McConville}, {McEnery}, {Mehault}, {Michelson}, {Minuti},
  {Mitthumsiri}, {Mizuno}, {Moiseev}, {Mongelli}, {Monte}, {Monzani},
  {Morselli}, {Moskalenko}, {Murgia}, {Nakamori}, {Naumann-Godo}, {Norris},
  {Nuss}, {Nymark}, {Ohno}, {Ohsugi}, {Okumura}, {Omodei}, {Orlando}, {Ormes},
  {Ozaki}, {Paneque}, {Panetta}, {Parent}, {Perkins}, {Pesce-Rollins},
  {Pierbattista}, {Pinchera}, {Piron}, {Pivato}, {Porter}, {Racusin},
  {Rain{\`o}}, {Rando}, {Razzano}, {Razzaque}, {Reimer}, {Reimer}, {Reposeur},
  {Ritz}, {Rochester}, {Romani}, {Roth}, {Rousseau}, {Ryde}, {Sadrozinski},
  {Salvetti}, {Sanchez}, {Saz Parkinson}, {Sbarra}, {Scargle}, {Schalk},
  {Sgr{\`o}}, {Shaw}, {Shrader}, {Siskind}, {Smith}, {Spandre}, {Spinelli},
  {Stephens}, {Strickman}, {Suson}, {Tajima}, {Takahashi}, {Takahashi},
  {Tanaka}, {Thayer}, {Thayer}, {Thompson}, {Tibaldo}, {Tibolla}, {Tinebra},
  {Tinivella}, {Torres}, {Tosti}, {Troja}, {Uchiyama}, {Vandenbroucke}, {Van
  Etten}, {Van Klaveren}, {Vasileiou}, {Vianello}, {Vitale}, {Waite},
  {Wallace}, {Wang}, {Werner}, {Winer}, {Wood}, {Wood}, {Wood}, {Yang}, \&
  {Zimmer}}]{FermiLATcatalog}
{Nolan}, P.~L., {Abdo}, A.~A., {Ackermann}, M., {$et~al$.} 2012, \apjs, 199, 31

\bibitem[{{Pletsch}(2010)}]{2010PhRvD..82d2002P}
{Pletsch}, H.~J. 2010, \prd, 82, 042002

\bibitem[{{Prix} \& {Shaltev}(2012)}]{2012PhRvD..85h4010P}
{Prix}, R., \& {Shaltev}, M. 2012, \prd, 85, 084010

\bibitem[{{Ransom} {$et~al$.}(2002){Ransom}, {Eikenberry}, \&
  {Middleditch}}]{ransom}
{Ransom}, S.~M., {Eikenberry}, S.~S., \& {Middleditch}, J. 2002, \aj, 124, 1788

\bibitem[{Rayleigh(1919)}]{Rayleigh1919}
Rayleigh, L. 1919, The London, Edinburgh, and Dublin Philosophical Magazine and
  Journal of Science, 37, 321

\bibitem[{{Singh} {$et~al$.}(2014){Singh}, {Tandon}, {Agrawal}, {Antia},
  {Manchanda}, {Yadav}, {Seetha}, {Ramadevi}, {Rao}, {Bhattacharya}, {Paul},
  {Sreekumar}, {Bhattacharyya}, {Stewart}, {Hutchings}, {Annapurni}, {Ghosh},
  {Murthy}, {Pati}, {Rao}, {Stalin}, {Girish}, {Sankarasubramanian},
  {Vadawale}, {Bhalerao}, {Dewangan}, {Dedhia}, {Hingar}, {Katoch}, {Kothare},
  {Mirza}, {Mukerjee}, {Shah}, {Shah}, {Mohan}, {Sangal}, {Nagabhusana},
  {Sriram}, {Malkar}, {Sreekumar}, {Abbey}, {Hansford}, {Beardmore}, {Sharma},
  {Murthy}, {Kulkarni}, {Meena}, {Babu}, \& {Postma}}]{2014_sing_AstroSat}
{Singh}, K.~P., {Tandon}, S.~N., {Agrawal}, P.~C., {$et~al$.} 2014, in
  \procspie, Vol. 9144, Space Telescopes and Instrumentation 2014: Ultraviolet
  to Gamma Ray, 91441S

\bibitem[{{Singhal} {$et~al$.}(2019){Singhal}, {Leaci}, {Astone}, {D'Antonio},
  {Frasca}, {Intini}, {La Rosa}, {Mastrogiovanni}, {Miller}, {Muciaccia},
  {Palomba}, \& {Piccinni}}]{2019CQGra..36t5015S}
{Singhal}, A., {Leaci}, P., {Astone}, P., {$et~al$.} 2019, Classical and
  Quantum Gravity, 36, 205015

\bibitem[{Virtanen {$et~al$.}(2020)Virtanen, Gommers, Oliphant, Haberland,
  Reddy, Cournapeau, Burovski, Peterson, Weckesser, Bright, {van der Walt},
  Brett, Wilson, Millman, Mayorov, Nelson, Jones, Kern, Larson, Carey, Polat,
  Feng, Moore, {VanderPlas}, Laxalde, Perktold, Cimrman, Henriksen, Quintero,
  Harris, Archibald, Ribeiro, Pedregosa, {van Mulbregt}, \& {SciPy 1.0
  Contributors}}]{2020SciPy-NMeth}
Virtanen, P., Gommers, R., Oliphant, T.~E., {$et~al$.} 2020, Nature Methods,
  17, 261

\end{thebibliography}



\appendix

\section{Behaviour of $Z_n^2(f)$}
\label{sec:fullformula}
Here, we delve into certain behaviours of the $Z_n^2(f)$ metric. While these observations primarily stem from our error analysis study, they illuminate specific statistical properties of the metric, further enriching our understanding of its characteristics and implications.

In our study, we emphasized the 
${Tb^2}/{2a}~\rm{sinc}^2\left[T\pi(f-f_0)\right]$ term, which is the dominant sinc-squared term around the source frequency. This is because other terms were numerically insignificant for reasonably longer observation times. The $Z_n^2$ function is presented as:

\begin{align}
Z_1^2(f) &= \frac{Tb^2}{2a}\rm{sinc}^2\left[T\pi(f-f_0)\right] + \frac{T(2a)^2}{2a}\rm{sinc}^2\left(\pi Tf\right)  \nonumber\\ 
&\quad+ \frac{Tb^2}{2a}\rm{sinc}^2\left[T\pi(f+f_0)\right] + \text{other terms} .\label{zn2sine2}
\end{align}

However, there exists an additional cross term, \begin{align}
    \frac{Tb^2}{a}\rm{sinc}\left[T\pi(f+f_0)\right]\rm{sinc}\left[T\pi(f-f_0)\right],\label{eq:crossterm}
\end{align} 
that we did not discuss. Theoretical calculations and simulations show that this term is numerically negligible around the source frequency. 

\begin{figure}
\centering
\includegraphics[width=1.0\linewidth]{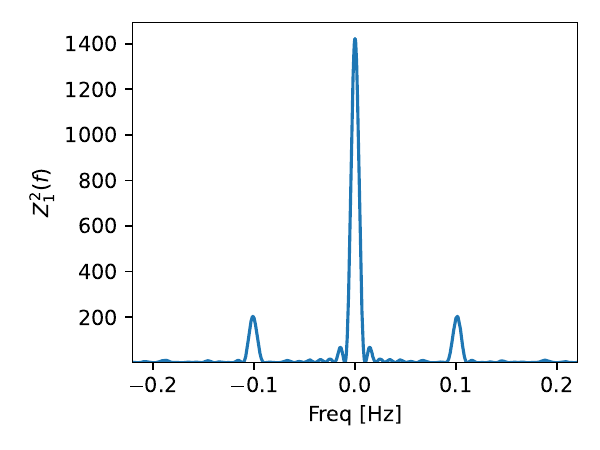}
\caption{Plot of $Z_1^2(f)$ for sinusoidal data, as delineated by Eq.~\ref{zn2sine}. As the data is real valued the plot is symmetric in negative frequencies. The figure emphasizes the minimal influence of the cross term around the source frequency, as described in Eq.~\ref{eq:crossterm}.}
\label{fig:completeformula}
\end{figure}

Utilizing both the developed algorithm and the capabilities of the \texttt{Stingray} software to accommodate negative frequencies, we can depict a sinusoidal signal comprehensively, as illustrated in Fig.~\ref{fig:completeformula} for a sinusoidal Poisson data characterized by $(a,b,f_0,T) = (7, 5, 0.1, 100)$. The figure portrays the sum of three sinc-squared functions, with an imperceptible contribution from the cross term. In many practical scenarios, these sinc-squared terms can be deemed distinct since they typically don't impinge upon each other's localized values. Nonetheless, this approximation tends to be less valid for lower signal source frequencies. Additionally, the presence of low-frequency noise, such as red noise, might perturb the analysis.

A closer examination of the cross term reveals that it does not necessarily peak at the source frequency. This characteristic might affect peak measurements, especially when the SNR is low. Although we have not delved deeply into this aspect, it seems that the curve fitting routine in its current form might not benefit from including this cross term. 

When considering higher harmonics, where the assumption of high SNR does not hold, the contribution in the $Z_n^2$ term can still be significant due to the $k^2$ term (with $k$ being the harmonics). In such scenarios, the $Z_n^2$ statistics deviate from the sinc-squared approximation.

\begin{figure}
\centering
\includegraphics[width=1.0\linewidth]{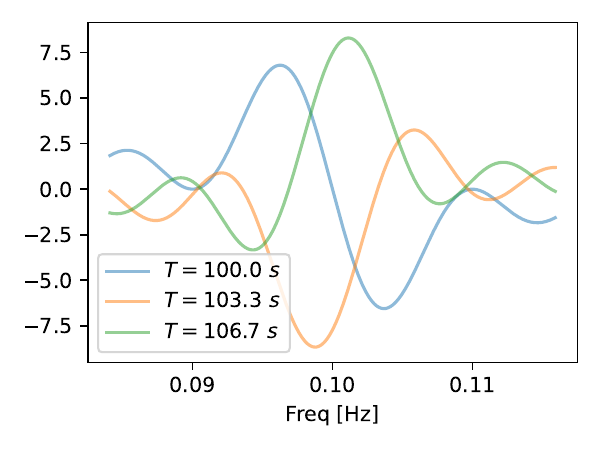}
\caption{A plot of the cross term in $Z_{1}^{2}$ statistics for the same source parameters as in Fig.~\ref{fig:completeformula}, with varied $T_{\mathrm{obs}}$. Although it is relatively insignificant in comparison to the local sinc-squared term, its first and second derivatives exhibit notable differences from those of the sinc-squared term.}
\label{fig:cross-term}
\end{figure}

\begin{figure}
\centering
\includegraphics[width=1.0\linewidth]{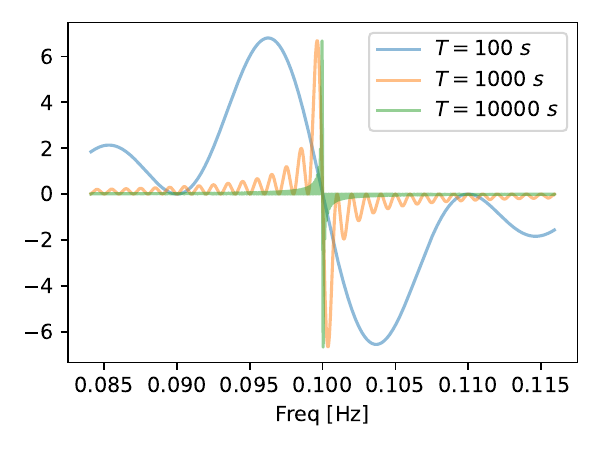}
\caption{{Comparison of the cross term over long observation times, $T_{\mathrm{obs}}$. As $T_{\mathrm{obs}}$ grows, the numerical impact of the cross term becomes significantly less noticeable.}
}
\label{fig:cross-time}
\end{figure}

However, this behavior diminishes when averaged out over a large number of realizations. The exact contribution of this term and the optimal scenario for its incorporation remain unclear and may be the subject of future studies. 

Figures \ref{fig:cross-term} and \ref{fig:cross-time} present a detailed examination of the cross terms. Compared to the sinc-square behavior depicted in Fig.~\ref{fig:completeformula}, it becomes evident that these cross terms carry a notable significance. We observe that the peak position and its maximum value display varies with $T_{\rm{obs}}$. However, despite these variations, it remains relatively modest, even for an extended observation duration. Although the numerical impact appears minimal, it remains to be determined whether this contributes to the statistical behavior influencing the measurement of the source frequency.

\subsection{Expectation and Variance of $Z_n^2(f)$} \label{subsec:expectation_variance}

In this section, we rigorously investigate the expectation and variance of the quantity $Z_n^2(f)$, which holds a central role in our analytical framework. Our analysis assumes the data to be observed over an extended period, characterized by a large observation time $T_{\text{obs}}$.

We commence by examining the expectation of $Z_n^2(f)$ under these conditions by considering,

\begin{align}
E\left[Z_{n}^2(f)\right] &= E\left[\frac{2}{N} \sum_{k=1}^{n} \left( \sum_{m=1}^{n_b} w_m\sin{(k\phi^{\prime}_m)} \right)^2 \right. \nonumber\\ 
&\left. \hspace{2cm} + \left( \sum_{m=1}^{n_b} w_m\cos{(k\phi^{\prime}_m)} \right)^2 \right].
\end{align}

In our analysis, the quantities $w_m$ represent the number of photons within a phase bin, following a Poisson distribution. An important property aiding this analysis is that Poisson distribution of a sum equals the sum of Poisson distributions. Formally, this property can be expressed as,

\begin{equation}
    \text{Poisson}(\lambda_i + \lambda_j) = \text{Poisson}(\lambda_i) + \text{Poisson}(\lambda_j).
\end{equation}

Considering extended observation times, where $w_m \gg 1$ and leveraging the equivalence between the expectation, variance, and the Poisson rate of a Poisson process, we deduce,

\begin{align}
    E\left[ w_i^2\right] &= E\left[ w_i\right]^2 + \rm{Var}\left[ w_i\right],\\
    &= E\left[ w_i\right]^2 + E\left[ w_i\right],\\
    &\approx E\left[ w_i\right]^2.
\end{align}

Additionally, since $w_i$ and $w_j$ are independent for $i \neq j$, we further simplify the analysis:

\begin{align}
    E\left[ w_iw_j\right] = E\left[ w_i\right]E\left[ w_j\right].
\end{align}

Combining these results, we derive the expectation of $Z_n^2(f)$ as,

\begin{align}
    E\left[Z_n^2(f)\right] &= \frac{2}{N} \sum_{k=1}^{n} \left( \sum_{m=1}^{n_b} w_m\sin{(k\phi^{\prime}_m)} \right)^2  \nonumber\\ 
& \hspace{2cm} + \left( \sum_{m=1}^{n_b} w_m\cos{(k\phi^{\prime}_m)} \right)^2,\\
&= Z_n^2(f).
\end{align}

Furthermore, to validate our theoretical derivation, we conduct numerical simulations. Specifically, we simulate several relaization of Poisson dominate sinusoidal signal and measure the mean and variance of $Z_n^2(f)$.

\subsubsection{Numerical computation of expectation and variance}
\begin{figure}
    \centering
    \includegraphics[width=1.0\columnwidth]{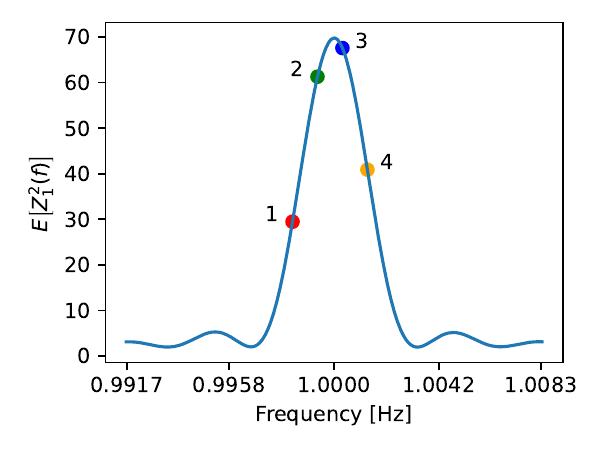}
    \caption{The plot shows the expected value of $Z_1^2(f)$ derived from simulated sinusoidal Poisson data, characterized by parameters $(a, b, f_0, T) = (6.5, 3.5, 1.0, 300)$ and employing $n_{\text{sim}} = 10,000$ simulations. Superimposed are four distinct mean values of $Z_1^2(f)$ corresponding to the frequencies $f_1$, $f_2$, $f_3$, and $f_4$, denoted by the colors red, green, blue, and orange, respectively.}

    \label{fig:zn2_scatter}
\end{figure}

\begin{figure}
    \centering
    \includegraphics[width=1.0\columnwidth]{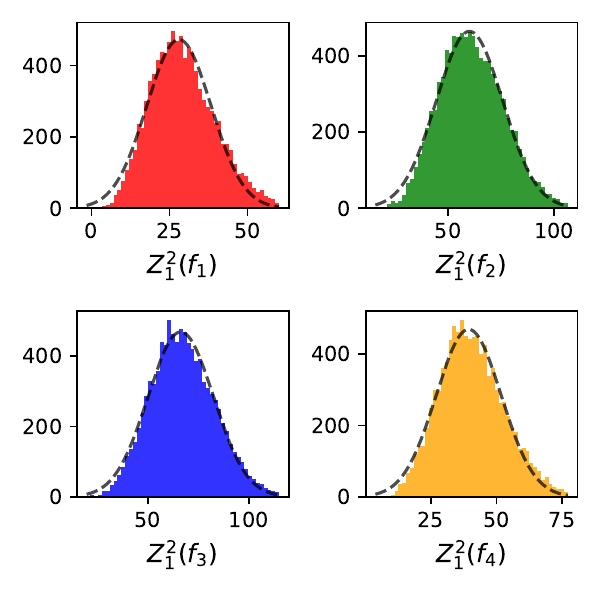}
    \caption{The figure presents a 2x2 grid of plots, each showcasing histograms of the $Z_1^2$ values at distinct frequencies ($f_1$, $f_2$, $f_3$, and $f_4$) from different realizations, each based on $n_{\text{sim}} = 10,000$ simulations. Specifically, the top-left plot corresponds to $f_1$, the top-right to $f_2$, the bottom-left to $f_3$, and the bottom-right to $f_4$. The colors of the histograms align with those used in the preceding figure, maintaining consistency in representation.}
    \label{fig:zn2_pdf}
\end{figure}

\begin{figure}
    \centering
    
    \begin{subfigure}[b]{\columnwidth}
        \centering
        \includegraphics[width=1.0\columnwidth]{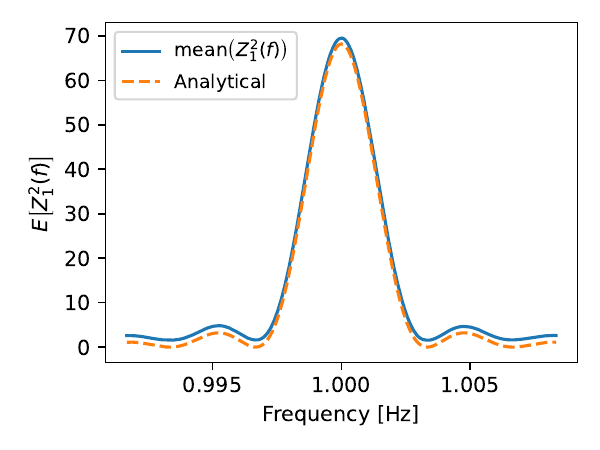}
        \caption{Empirical and analytical comparison of the expectation of $Z_1^2(f)$ for sinusoidal Poisson data. The analytical sinc-squared function is given by $\frac{Tb^2}{2a}~\rm{sinc}^2\left[T\pi(f-f_0)\right]$, which exhibits excellent agreement with the data.}
        \label{fig:mean_analy_z12}
    \end{subfigure}
    
    \vspace{0.2cm}  
    
    \begin{subfigure}[b]{\columnwidth}
        \centering
        \includegraphics[width=1.0\columnwidth]{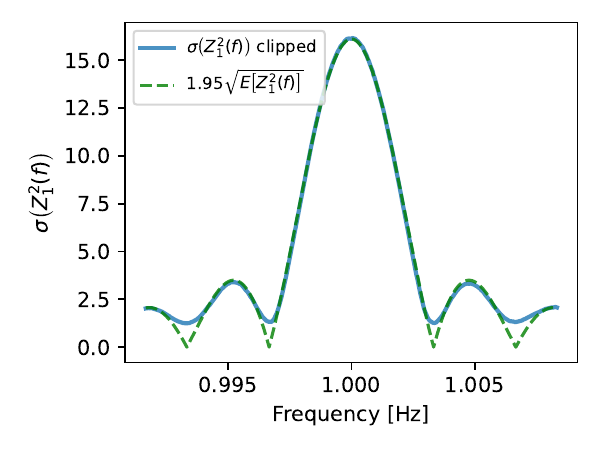}
        \caption{Standard variance of $Z_1^2(f)$ plotted as a function of $f$. As indicated in Fig.~\ref{fig:zn2_pdf}, empirically, it approximates to $2 \sqrt{E\left[Z_1^2(f)\right]}$.}
        \label{fig:std_analy_z12}
    \end{subfigure}

    \caption{An empirical comparison of the expectation and standard deviation of $Z_1^2(f)$ shown in \textit{top} and \textit{bottom} plot respectively.}
    \label{fig:analy_z12}
\end{figure}

Since the noise is correlated in Poisson limited data, deriving \text{Var}$[Z_n^2(f)]$ becomes challenging. However, using simulations, we validate the above assertion regarding the expectation of $Z_n^2(f)$ and empirically investigate the relationship of $\text{Var}[Z_n^2(f)]$ as presented.

For sinusoidal Poisson data characterized by the parameters $(a,b,f_0,T) = (100, 90, 1, 300)$, we systematically compute $Z_n^2(f)$ across multiple realizations for a consistent set of frequencies. By creating a histogram for the values of $Z_1^2(f_i)$ computed for each distinct $f_i$, we can determine both the expectation and variance associated with each $Z_1^2(f_i)$ at the respective frequency $f_i$. Figures \ref{fig:zn2_scatter} and \ref{fig:zn2_pdf} display these findings. To extract meaningful statistical insights from the histogram, we fit it using a Gaussian distribution, enabling us to ascertain the mean and standard deviation. By juxtaposing our obtained results with theoretical predictions, we are able to empirically derive and validate the governing function behind this behavior.

\begin{figure}
    \centering
    \includegraphics[width=1.0\columnwidth]{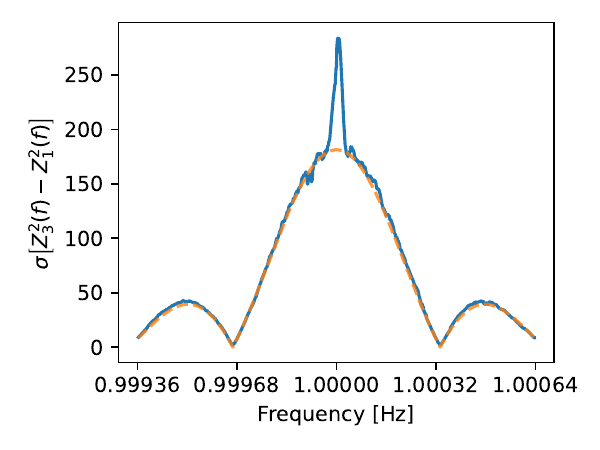}
    \caption{Standard deviation of the difference $Z_3^2(f)-Z_2^2(f)$ for rectangular signal data. A prominent peak is observed, diverging from the anticipated empirical shape.}

    \label{fig:my_label} 
\end{figure}
However, we also highlight that for higher harmonics, there exists a pronounced peak-like fluctuation in the $sinc^2$-like behavior for $Z_n^2(f) - Z_{n-1}^2(f)$, evident in its variance. This pattern, though not extensively studied, has been consistently observed in higher harmonics across a broad spectrum of source parameters.
Nevertheless, these fluctuations near the peak introduce additional complications, which we address in the subsequent subsection.

\subsection{Maxima vs Curve Fit}\label{max-cur}
As $Z_n^2(f)$ approaches its maximum, the relative fluctuation amplifies, and in higher harmonics, the fluctuation is substantially more pronounced. Often, the peak of $Z_n^2(f)$ is utilized as a measure of frequency and associated parameters. Consequently, fluctuations near this peak can introduce errors, particularly when deriving the frequency from the maximum value. Through simulations, we further demonstrate that the frequency variance, when determined from the peak, is considerably higher than when derived via curve fitting. Furthermore, the curve fitting method yields results more aligned with anticipated values, as depicted.

\begin{figure}
    \centering
    \includegraphics[width=1.0\columnwidth]{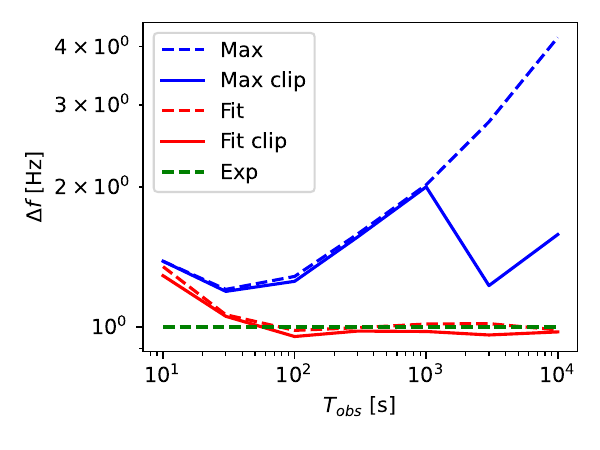}
    \caption{Comparison of frequency estimation uncertainty between the 'maximum of $Z_1^2(f)$' (blue) method and the curve fitting (red) approach. Both methods competently estimate the signal frequency however, the uncertainty associated with the 'maximum' method is several times higher. Although it depends on parameter space, despite applying a three-sigma ($3\sigma$) clipping to reduce outliers, the standard deviation of the 'maximum' method's estimates remains twice as large as that of both the curve fitting method and the analytical formulation (green). The simulations are based on sinusoidal signal with parameters $(a,b,f_0,T) = (100,30,1,200)$}
    \label{fig:max_fit}
\end{figure}

\subsection{Fine tuning the curve fitting process}\label{sec:finetune}
Given our observations of notable fluctuations around the peak, it is clear that estimating measured quantities by relying solely on the maxima can introduce significant errors into calculations. While our method primarily hinges on curve fitting, the range of our fitting data is still determined by the maxima's location, a practice that may not always be optimal and could introduce errors, particularly with higher harmonics.

We have explored several methods to enhance the robustness and reliability of this approach. A preliminary understanding based on simulations and results suggests a new strategy: instead of using a centered range defined by $ {1}/{(\pi T_{\rm{obs}})} $ on each side, we identify the full-width at half maximum (FWHM) based on the maxima. This approach yields a more symmetric data set for curve fitting and reduces the risk of incorporating noise present in the data tails. Moreover, given the sharp peak observed in a narrow range around the maxima, we suggest excluding these data prior to curve fitting. Preliminary visual assessments indicate that this method produces superior curve-fitting results.

Nonetheless, a comprehensive study verifying the robustness and reliability of these methods across the entire spectrum is yet to be conducted. Our current insights are preliminary and are based on selective simulations and results.
Furthermore, while we utilize a region of width $ w $ around the peak of $ Z_n^2 $ as previously detailed, our studies have highlighted the following intricacies: the narrower the region around the measured frequency used for curve fitting, the closer its behavior resembles the sinc-squared function. Nonetheless, caution must be exercised. Selecting an exceedingly narrow region might ensure a sinc-squared behavior, but this comes at the cost of increased noise, resulting in a less-than-optimal signal-to-noise ratio for reliable curve fitting. This introduces a trade-off regarding the optimal size of the curve fitting region.

Referring to Eq.~\ref{abt2}, the region may be less than an order of $\sim  {1}/{T_{\rm{obs}}^{5/4}} $ to diminish the high order terms in Taylor expansion of $Z_1^2(f)$. It's worth noting that these insights, while valuable, are based on initial findings and were not incorporated into our primary robust method. A comprehensive comparison remains to be undertaken. Future efforts will likely fine-tune these methodologies to yield more accurate curve fitting. This optimization can further alleviate potential errors in both frequency and error estimations resulting from suboptimal curve fitting.

\subsection{Time gaps}\label{sec:time--gaps}
\begin{figure*}
    \centering
    \includegraphics[width=\textwidth]{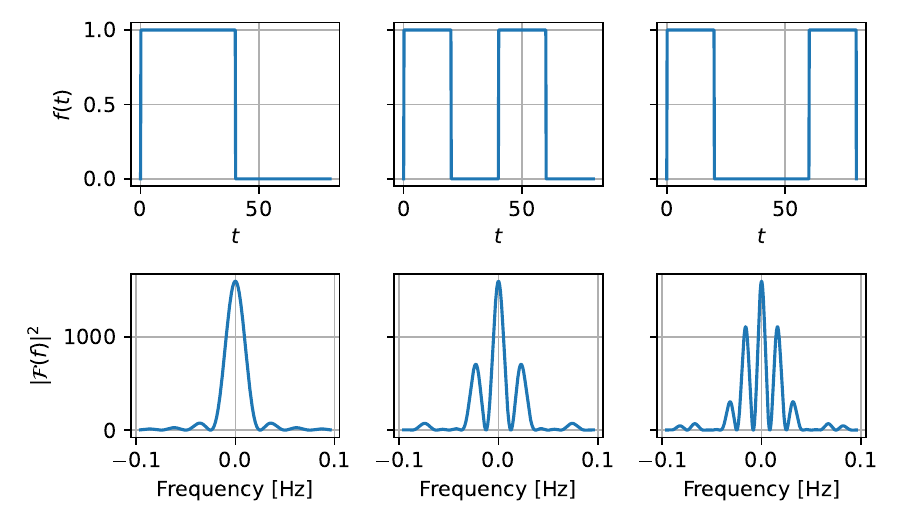}
    \caption{For signal data with gaps, $Z_n^2(f)$ deviates from the sinc-squared function and is better described as a sum of rectangular functions. By computing the Fourier transform of this summed representation, we can approximate $Z_n^2(f)$ for gapped data. This figure contrasts the absolute Fourier values with the expected $Z_n^2(f)$ of a sinusoid bearing identical gaps.}
    \label{fig:time_gap}
\end{figure*}

\begin{figure*}
    \centering
    \includegraphics[width=\textwidth]{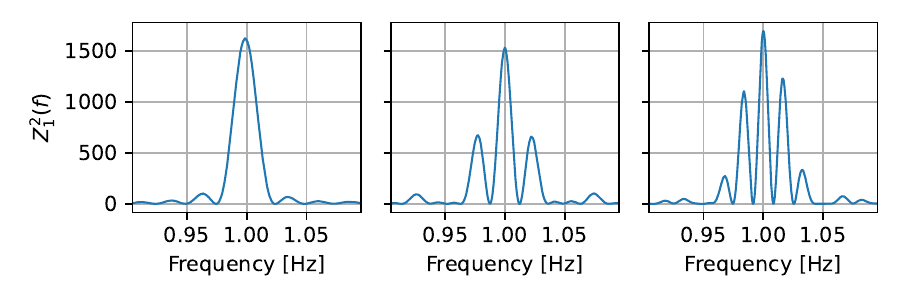}
    \caption{For sinusoidal Poisson data simulated with parameters $(a, b, f, T) = 100,90,1, 40$, with time gaps as 0, 20, 40 sec respectively, which is same as depicted in the prior figure.  The $Z_1^2(f)$ of one such realization aligns closely with the Fourier transform of the rectangular window function. This alignment is also visually consistent with the approximated $Z_n^2(f)$ formula given in Eq.~\ref{gap_z12}.}
    \label{fig:time_gap_Z12}
\end{figure*}

Consider a Poisson data where the source is observed during specific intervals and the data are either removed or unobserved during others. Let us denote these intervals by $T_n$, where odd-indexed $T$ values (\textit{e.g.,} $T_1, T_3, T_5, \ldots$) correspond to the observed periods, and even-indexed values (\textit{e.g.,} $T_2, T_4, T_6, \ldots$) correspond to unobserved or removed time intervals.

For each observed interval $T_i$, the signal contribution can be defined as:
\begin{equation}
s_j(f) = T_j \exp\left(-i \left[\sum_{k=1}^{i} T_k - \frac{T_j}{2} \right] 2\pi f \right) \text{sinc}\left[T_j (f - f_0)\right].
\end{equation}

Here, $f$ is the frequency of interest, and $f_0$ denotes a reference frequency. The exponential term captures the phase shift due to the time elapsed, and the sinc function models the effect of the windowing in frequency space.

Thus, the $Z_1^2(f)$, considering all observed intervals, is the norm squared of summation of $s_j(f)$ as,
\begin{equation}
Z_1^2(f) \approx \left|\sum_{i \in \text{odd}} s_i(f)\right|^2. \label{gap_z12}
\end{equation}

Utilizing the derived formulation, we can compute the second derivative, a crucial component for error estimation. The expression is given by,
\begin{align}
    \cfrac{b^2\pi^2}{6a\left(\sum_{m \in \rm{odd}}T_m\right)}\sum_{i=0}^{n-1}\left(-1\right)^{i+1}\sum_{j=1}^{n-i}\left(\sum_{k=0}^{i}T_{j+k}\right)^4.
\end{align}
Interestingly, when $T_{i>1} = 0$, this equation reverts to its original form. Furthermore, it's worth noting that while many pipelines employ the \textit{gti} for data analysis, facilitating the computation of $T_j$, it is not advisable to incorporate this large number of parameters into the curve fitting routine in its current form. 

Figs.~\ref{fig:time_gap} and~\ref{fig:time_gap_Z12} provide empirical validation for the model described in Eq.~\ref{gap_z12}. Fig.~\ref{fig:time_gap} shows the Fourier transforms of observation windows, or \textit{gti}, for sinusoidal Poisson data. In these windows, a value of one indicates periods of observation, while zero signifies periods when the source was not observed or data was removed. This is essential to understand how the gaps in observation impact the Fourier spectrum, affecting the detection of frequency in a sinusoidal signal, as elaborated in Eq.~\ref{FT}.

Fig.~\ref{fig:time_gap_Z12} showcases the simulation results of the $Z_1^2(f)$ statistic for sinusoidal data with parameters $(a,b,f_0,T) = (100, 90, 1, 40)$. The simulation varies $T_2$ within the set $\{0,20,40\}$, keeping $(T_1, T_3)$ constant at $(20,20)$. This approach evaluates the influence of different observation windows on the $Z_1^2(f)$ statistic, crucial for accurate signal frequency detection in astrophysical data analysis.

The results from these figures closely align with the analytical predictions in Section~\ref{sec_analytical}, substantiating the accuracy of the extended equation. The reliability of the $Z_1^2(f)$ method is evident, even with irregular observation windows. The good agreement between the simulation findings and the theoretical model underscores the importance of detailed analysis of the window function for accurate frequency estimation in astrophysics, especially when data irregularities are present.

The analytical modeling of $Z_n^2$ statistics for data with gaps, along with the statistical insights discussed, provides a clearer and more reliable understanding of how various factors influence the recovery of a source's frequency and its associated uncertainty. This knowledge is crucial for effectively modeling periodic high-energy sources or any Poisson-dominated data. Although our current approach prioritizes simplicity and robustness, and does not delve into all these aspects in depth. Future research could refine and extend this methodology, enhancing the precision of frequency estimation in astrophysical studies.

\end{document}